\documentclass[aps,pre,reprint,superscriptaddress]{revtex4-1}

\pdfoutput=1

\usepackage{graphicx}
\usepackage{amssymb}
\usepackage{amsmath}
\usepackage{xcolor}
\usepackage{bm} 
\usepackage{pdfpages}
\usepackage{pgffor}
\usepackage{CJK}
\usepackage{multirow}

\makeatletter
\AtBeginDocument{\let\LS@rot\@undefined}
\makeatother

\begin{document}

\title{Coarse-Grained Molecular Dynamics Modeling of a Branched Polyetherimide} 

\author{Chengyuan Wen}
\affiliation{Department of Physics, Center for Soft Matter and Biological Physics, and Macromolecules Innovation Institute, Virginia Polytechnic Institute and State University, Blacksburg, Virginia 24061, USA}
\author{Roy Odle}
\affiliation{SABIC, 1 Lexan Lane, Mt. Vernon, Indiana 47620, USA}
\author{Shengfeng Cheng}
\email{chengsf@vt.edu}
\affiliation{Department of Physics, Center for Soft Matter and Biological Physics, and Macromolecules Innovation Institute, Virginia Polytechnic Institute and State University, Blacksburg, Virginia 24061, USA}
\affiliation{Department of Mechanical Engineering, Virginia Polytechnic Institute and State University, Blacksburg, Virginia 24061, USA}

\date{\today}

\begin{abstract}
A coarse-grained model is developed to allow large-scale molecular dynamics (MD) simulations of a branched polyetherimide derived from two backbone monomers [4,4'-bisphenol A dianhydride (BPADA) and m-phenylenediamine (MPD)], a chain terminator [phthalic anhydride (PA)], and a branching agent [tris[4-(4-aminophenoxy)phenyl] ethane (TAPE)]. An atomistic model is first built for the branched polyetherimide. A systematic protocol based on chemistry-informed grouping of atoms, derivation of bond and angle interactions by direct Boltzmann inversion, and parameterization of nonbonded interactions by potential of mean force (PMF) calculations via gas-phase MD simulations of atomic group pairs, is used to construct the coarse-grained model. A six-pair geometry, with one atomic group at the center and six replicates of the other atomic group placed surrounding the central group in a NaCl structure, has been demonstrated to significantly speed up the PMF calculations and partially capture the many-body aspect of the PMFs. Furthermore, we propose a correction term to the PMFs that can make the resulting coarse-grained model transferable temperature-wise, by enabling the model to capture the thermal expansion property of the polymer. The coarse-grained model has been applied to explore the mechanical, structural, and rheological properties of the branched polyetherimide.
\end{abstract}

\pacs{}

\maketitle 

\section{Introduction}

Polyetherimides, as a special class of polyimide polymers, are the products of condensation reactions among bifunctional carboxylic anhydrides containing ether linkages and primary diamines.\cite{Urakawa2014} They are an important type of engineering thermoplastics with broad applications in industry, agriculture, transport, and healthcare because of their high heat resistance and stability, high mechanical strength, excellent electrical properties over a wide range of temperatures and frequencies, improved melt processability, good adhesive properties, and good chemical and environmental stability.\cite{Urakawa2014, johnson1983, Serfaty1984, Seifert2002, McKeen2017Book, MaYang2019} For example, Ultem is a family of polyetherimide products derived from the BPADA dianhydride and MPD diamine\cite{note_chem_formula_all} that has superior heat, solvent, and flame resistance and has been widely used in the automotive industry, medical and chemical instrumentation, and aerospace engineering.\cite{white1981, johnson1983, wright1991} To simplify the discussion, all abbreviations and acronyms of chemical formulae used in this paper are summarized in Ref.~[\onlinecite{note_chem_formula_all}]. Since the development of Ultem, there has been a strong interest in discovering and synthesizing new polyetherimides that possess improved existing or desirable new properties and functions.\cite{vora2000synthesis, tawade2015synthesis, rigana2016synthesis, kaya2018synthesis, cao2019, cao2020a, cao2020b, dennis2020} The Edisonian approach of trial and error is of course possible but becomes expensive and time-consuming when there is a need to scan a wide range of potential chemical formulae. To expedite materials discovery, computational approaches including molecular dynamics (MD) simulations have evolved into indispensable tools.\cite{Audus2017}

Over the past two decades, all-atom MD simulations with various force fields have been applied to investigate polyetherimides.\cite{YoungHinkley1999, QiHinkleyHe2005, Eichinger2002, Lyulin2013, Nazarychev2013, Lyulin2014, Larin2014, Falkovich2014a, Falkovich2014b, Falkovich2016, Nazarychev2016a, Nazarychev2016b, Nazarychev2017, Minelli2012, Lim2003, Lim2007, Luchinsky2018, Luchinsky2019, Pinel2002, Neyertz2006, Neyertz2007, Neyertz2008, Neyertz2009, Pan2007, Pan2010, Xia2010, ZhangSundararaj2003, Zhao2018, Wen2020, deNicola2017, Hamm2017} Young \textit{et al.} computed the dielectric relaxation strength of polyetherimides based on the ODPA dianhydride and APB diamine or its nitrile substituted version using atomistic MD simulations.\cite{YoungHinkley1999} Qi \textit{et al.} simulated a composite of carbon nanotubes (CNTs) and the LARC-SI polyimide, which is a copolymer based on two dianhydrides (ODPA and BPDA) and the 3,4'-ODA diamine, and computed its glass transition temperature, and stress-strain curves, Young's moduli, densities, and Poisson ratios at various temperatures.\cite{QiHinkleyHe2005} Eichinger \textit{et al.} computed the solubility parameter of Ultem and evaluated the interfacial interactions between Ultem and a variety of low molecular-weight liquids.\cite{Eichinger2002} In a series of papers, Lyulin and collaborators have performed all-atom MD simulations on a range of polytherimides, including those based on the dianhydride R and various diamines (e.g., BAPS, BAPO, and BAPB),\cite{Lyulin2013, Nazarychev2013, Lyulin2014, Larin2014, Falkovich2016, Nazarychev2016a, Nazarychev2016b}, Ultem and Extem (a polyetherimide based on the BPADA dianhydride and DDS diamine)\cite{Falkovich2014a, Falkovich2014b}, and the ODPA-P3 polyetherimide and BPDA-P3 and aBPDA-P3 polyimides based on the P3 diamine.\cite{Nazarychev2017} They have demonstrated a two-step protocol that allows microsecond all-atom MD simulations for polymer equilibration,\cite{Lyulin2013, Nazarychev2013} computed the thermal properties of bulk R-BAPS and Extem,\cite{Lyulin2014} and identified the ordering behavior of R-BAPS and R-BAPB at the surface of a single-walled CNT\cite{Larin2014} and a graphene sheet.\cite{Falkovich2016} They also investigated the influence of electrostatic interactions on the thermophysical properties of Ultem and Extem,\cite{Falkovich2014a, Falkovich2014b} and R-BAPS.\cite{Nazarychev2016a} Furthermore, they simulated the mechanical deformation and computed the elastical moduli of various polyimides.\cite{Nazarychev2016b, Nazarychev2017}

Minelli combined MD simulations and perturbed-chain statistical associating fluid theory to predict the solubility of various gases in Ultem and Kapton (a PMDA-ODA polyimide).\cite{Minelli2012} Lim \textit{et al.} simulated the diffusion and sorption of $\text{CO}_2$ and $\text{CH}_4$ in amorphous Ultem and a CNT-Ultem composite.\cite{Lim2003, Lim2007} Luchinsky \textit{et al.} computed the vibrational and infrared spectra and the thermodynamic and rheological properties of a mixture of Ultem and polycarbonate via all-atom MD simulations.\cite{Luchinsky2018} They also investigated the diffusion and reptation dynamics of polymer chains at a polymer-polymer interface in the mixture.\cite{Luchinsky2019} Pinel, Brown, Neyertz \textit{et al.} used MD simulations to probe the effect of the rigidity of the dianhydride monomer (ODPA versus BCDA) and the addition of trifluoromethyl or methoxy substituents to the diamine monomer on the structure of the ODPA-ODA-$b$-BCDA-ODA copolyimides.\cite{Pinel2002} Subsequently, Neyertz, Brown, and collaborators simulated the permeation and sorption of various gases, including helium and oxygen, in polyimides and copolyimides based on the ODPA and/or BCDA dianhydride and ODA diamine.\cite{Neyertz2006, Neyertz2007, Neyertz2008, Neyertz2009} Pan \textit{et al.} simulated the ODPA-MPD-$b$-ODPA-ODA copolyimides with biphenyl side chains, which contains different numbers of methylene spacing groups, attached to the MPD monomers.\cite{Pan2007} They further applied a similar simulation model to investigate how the length of the alkyl side chains tethered to the MPD monomers influences the properties of Ultem.\cite{Pan2010} Xia \textit{et al.} computed the fractional accessible volume of Extem, Ultem, and polysulfone.\cite{Xia2010} Zhang \textit{et al.} used MD simulations to study the miscibility of Ultem and polycarbonates.\cite{ZhangSundararaj2003} Zhao \textit{et al.} employed MD simulations to reveal the effects of various silane coupling agents on the interface between silica and a ODPA-ODA polyetherimide.\cite{Zhao2018} Wen \textit{et al.} computed the glass transition temperature of various polyetherimides based on the BPADA dianhydride and different diamines with all-atom MD simulations and derived a predictive model of their glass transition temperatures using machine-learning algorithms.\cite{Wen2020} De Nicola \textit{et al.} combined experiment and all-atom MD simulations to investigate the local structure and dynamics of water absorbed in Ultem.\cite{deNicola2017} Additionally, Hamm \textit{et al.} used the ReaxFF reactive force field to model the pyrolysis process of Ultem.\cite{Hamm2017}

All-atom MD simulations have significantly deepened our understanding of polyimides and polyetherimides at the molecular level. However, they are still limited to relatively small systems and short time scales.\cite{Lyulin2013, Barrat2010} To access larger size and longer time scales, we have to resort to other computational techniques. One approach is to employ a coarse-grained description of a polymer system to get rid of fast degrees of freedom that play a less important role in the physical properties and processes of interest.\cite{Noid2013, Guenza2017} The typical practice, called ``coarse-graining'', is to group atoms into coarse-grained beads and parameterize the bead-bead interactions on the basis of all-atom simulations or available experimental data on material properties.\cite{Noid2013, Guenza2017}

Several methods of deriving coarse-grained potentials, including bottom-up approaches such as iterative Boltzmann inversion (IBI),\cite{Soper1996, Reith2003, Moore2014} inverse Monte Carlo,\cite{Lyubartsev1995, Lyubartsev1997} force matching,\cite{Ercolessi1994, Izvekov2005JCP, Izvekov2005JPCB, Noid2008a, Noid2008b, Wang2009}, the relative entropy,\cite{Shell2008, ChaimovichShell2011} the reversible work,\cite{McCoy1998, brini2011conditional}, energy renormalization,\cite{Xia2017ER, Xia2019SciAdv} and various top-down or mixed approaches,\cite{Shelley2001, Shinoda2007, Potter2019, Huang2019} have been developed. Several groups have applied these methods to construct coarse-grained MD models of polyimides\cite{Clancy2004a, Clancy2004b, Odegard2005, Pandiyan2015, Kumar2018, Sudarkodi2018, HuLuGuo2019} and polyetherimides.\cite{Chakrabarty2010, Markina2017, Volgin2018} Clancy \textit{et al.} coarse-grained polyimides based on the BPDA dianhydride and three different APB diamine isomers into chains of linked vectors and used such linked-vector chains to quickly obtain relaxed configurations of the polymers that can be reverse-mapped to atomistic systems.\cite{Clancy2004a, Clancy2004b} Odegard \textit{et al.} used this technique to build all-atom configurations for the representative volume elements of the silica nanoparticle/BPDA-APB polyimide composites with various interfacial treatments.\cite{Odegard2005} Pandiyan \textit{et al.} used the IBI method to construct coarse-grained models of a high-temperature polyimide (HFPE-30) based on the 6FDA dianhydride and PPD diamine, terminated by the 4-PEPA anhydride, at different levels of detail.\cite{Pandiyan2015} Kumar \textit{et al.} constructed a coarse-grained model of PMR-15, which is a polyimide based on the BTDA dianhydride and MDA diamine with the NA anhydride as the chain terminator, and studied its adhesion property on a rutile surface.\cite{Kumar2018} Sudarkodi \textit{et al.} further used these types of coarse-grained models to simulate the uniaxial tensile deformation of PMR-15 and HFPE-52 polyimides.\cite{Sudarkodi2018}

A few coarse-graining attempts have also been reported on polyetherimides.\cite{Chakrabarty2010, Markina2017, Volgin2018} Chakrabarty and Cagin developed an atomistically informed coarse-grained model of a piezoelectric polyetherimide based on the ($\beta$-CN)APB diamine and ODPA dianhydride and used this model to study the thermal, mechanical, and electrical properties of the polymer.\cite{Chakrabarty2010} Markina \textit{et al.} developed coarse-grained models of two polyetherimides based on the dianhydride R and BAPS or BPAB diamine using the dissipative particle dynamics method and showed that the chain stiffness has a profound influence on their crystallization behavior.\cite{Markina2017} Volgin \textit{et al.} developed two coarse-gained models of a polyetherimide based on the dianhydride R and BAPS diamine using the IBI method and studied the effects of coarse-graining level on the diffusion of a $\text{C}_{60}$ nanoparticle in the polymer matrix.\cite{Volgin2018}

A typical challenge faced by most coarse-grained models of molecular systems is that the models are usually thermodynamic state-dependent and not transferable.\cite{Noid2013} Recently, Hu \textit{et al.} constructed a transferable coarse-grained MD model based on the IBI approach combined with the density correction method to predict the thermodynamic, structural, and mechanical properties of Kapton.\cite{HuLuGuo2019} Clearly, more efforts are needed along this direction for polyimide and polyetherimide polymers. In this paper, we aim to develop a coarse-grained model that is not only transferable but also expandable for a branched polyetherimide derived from Ultem. Being transferable means that the model can be used at different thermodynamic state points, such as different temperatures. The expandability requirement indicates that when a new functional group (e.g., a side group or an ionic group) is added to the polyetherimide for which a coarse-grained model is already available, we just need to expand the existing model by adding new beads corresponding to the newly added atomic groups. The task then becomes parameterizing the interactions between these new beads and the existing beads. We demonstrate in this paper that by combining a chemistry-informed grouping method, potential of mean force (PMF) calculations, and a correction to the PMFs, it is possible to construct a transferable and expandable coarse-grained model for the branched polyetherimide based on the BPADA dianhydride and MPD diamine, with the PA anhydride as a chain terminator and TAPE as a branching agent.

This paper is organized as follows. The general theory of the systematic force-matching coarse-graining method, first discussed by Voth, Andersen, Noid and coworkers,\cite{Izvekov2005JCP, Izvekov2005JPCB, Noid2008a, Noid2008b, Wang2009} and how it can be practically approximated and implemented, are introduced in Sec.~\ref{sec:CG_theory}. Then we apply this theory to develop a coarse-grained model of the branched polyetherimide, as detailed in Sec.~\ref{sec:CG_model_development}. In Sec.~\ref{sec:CG_model_application}, the coarse-grained model is applied to compute mechanical, structural, and rheological properties of the branched polyetherimide and the results are compared to those from all-atom MD simulations and experiments. Finally, conclusions are included in Sec.~\ref{sec:conclusion}.

\section{General Theory of Coarse-graining} \label{sec:CG_theory}

In this paper, the term ``coarse-graining'' refers to the process of deriving a coarse-grained description of a polymeric system from a fine-grained, atomistic model of the same system. In other words, atoms in the detailed description are grouped into coarse-grained beads (i.e., pseudo-atoms) with some fine details smoothed over. The coarse-graining procedure thus involves two key steps: the mapping from groups of atoms to coarse-grained beads and the determination of interactions between those beads. We are motivated by the consideration that the final coarse-grained model being developed for the prototypical branched polyetherimide should be transferable, reusable, and expandable when a new formula of polyetherimide is the target of modeling.

We will delay the discussion on the coarse-graining mapping to Sec.\ref{sec:CG_mapping}, where our coarse-grained model of the branched polyetherimide is introduced. In this section, we introduce the general theory underlying the force-matching coarse-graining approach adopted here to determine the coarse-grained force field (i.e., the interaction potentials between the coarse-grained beads), which was originally proposed by Voth, Andersen, Noid, and coworkers.\cite{Izvekov2005JCP, Izvekov2005JPCB, Noid2008a, Noid2008b, Wang2009} We then discuss the approximate implementation of this theory through PMF calculations for atomic groups. Technical issues in the implementation, including the many-body effects on the PMFs and how to fix the center of mass of an atomic group in the PMF calculations, are discussed in detail. Finally, we present a test case with a pair of benzene molecules to validate the coarse-graining approach.

\subsection{Theory of Voth, Andersen, Noid, and Coworkers}

We first briefly summarize the theory of multiscale coarse-graining formulated by Voth, Andersen, Noid, and coworkers.\cite{Noid2008a} Our goal is to map an atomistic system with $n$ atoms to a system of $N$ coarse-grained beads through a mapping function $\bm{M}_{\bm{R}_I}(\bm{r}^n)$ for coordinates and $\bm{M}_{\bm{P}_I}(\bm{p}^n)$ for momenta, where $\bm{R}_I$ and $\bm{P}_I$ are the coordinate and momentum of the $I$-th coarse-grained bead while $\bm{r}_i$ and $\bm{p}_i$ are atomistic coordinates and momenta. Usually, linear maps are used such as
 \begin{equation}\label{eq:CG_mapping_coord}
 \bm{M}_{\bm{R}_I}(\bm{r}^n) = \sum_{i=1}^{n}c_{Ii}\bm{r}_i~,
 \end{equation} 
and
 \begin{equation}\label{eq:CG_mapping_momentum}
 \bm{M}_{\bm{P}_I}(\bm{p}^n) = M_I\sum_{i=1}^n c_{Ii}\bm{p}_i/m_i~,
 \end{equation}
where $I$ runs from 1 to $N$, the notation $\bm{r}^n$ ($\bm{p}^n$) represents the collection of atomistic coordinates (momenta), $c_{Ii}$ are coefficients in the linear map, $m_i$ is the mass of the $i$-th atom, and $M_I$ is the mass of the $I$-th coarse-grained bead.

The Hamiltonian of the atomistic  system is 
\begin{equation}
h(\bm{r}^n,\bm{p}^n)=\sum_{i=1}^n\frac{1}{2m_i}\bm{p_i^2}+u(\bm{r}^n)~,
\end{equation}
where $u(\bm{r}^n)$ is the potential energy of the atomistic system. The coarse-grained Hamiltonian is 
\begin{equation}
H(\bm{R}^N,\bm{P}^N)=\sum_{I=1}^N\frac{1}{2M_I}\bm{P}^2_I+U(\bm{R}^N)~,
\end{equation}
where $\bm{R}^N$ ($\bm{P}^N$) represents the collection of coarse-grained coordinates (momenta) and $U(\bm{R}^N)$ is the potential energy of the coarse-grained system. The main task of coarse-graining is therefore to determine $U(\bm{R}^N)$.

For a perfect coarse-grained model, which yields identical equilibrium properties as the atomistic model, the coarse-grained potential can be written as \cite{Noid2008a}
\begin{align}
    U(\bm{R}^N) = -k_\text{B}T\ln z(\bm{R}^N)+\text{const}~,
\end{align}
where 
\begin{align}
z(\bm{R}^N) \equiv \int d\bm{r}^n \text{exp}\left(-\frac{u(\bm{r}^n)}{k_\text{B}T}\right)
\delta\left(\bm{M}_{\bm{R}^N}(\bm{r}^n)-\bm{R}^N\right)~.
\end{align}
Here $\delta \left( \bm{x} \right)$ is the Dirac delta function with a vector argument. The coarse-grained force field, $\bm{F}_I(\bm{R}^N) \equiv -\frac{\partial U(\bm{R}^N)}{\partial \bm{R}_I}$, is then given by
\begin{align}\label{eq:CG_force_field}
    \bm{F}_I(\bm{R}^N) =&\frac{k_\text{B}T}{z(\bm{R}^N)}\int d\bm{r}^n \text{exp}\left(-\frac{u(\bm{r}^n)}{k_\text{B}T}\right) \nonumber\\
    &\times \prod_{J\neq I}\delta\left(\bm{M}_{\bm{R}_J}(\bm{r}^n)-\bm{R}_J\right) \nonumber\\
    &\frac{\partial}{\partial \bm{R}_I}\delta\left(\sum_{i\in \mathcal{I}_I } c_{Ii}\bm{r}_i-\bm{R}_I\right)~,
\end{align}
where the set $\mathcal{I}_I$ consists of the indices of the group of atoms that is mapped into the $I$-th coarse-grained bead. For the partial derivative, the following identity holds,
\small
\begin{align}\label{eq:partial_derivative_transform}
\frac{\partial}{\partial \bm{R}_I}\delta\left(\sum_{i\in \mathcal{I}_I } c_{Ii}\bm{r}_i-\bm{R}_I\right) = - \frac{\partial}{c_{Ik}\partial \bm{r}_{k}}\delta\left( \sum_{i\in \mathcal{I}_I} c_{Ii}\bm{r}_i-\bm{R}_I \right)~,
\end{align}
\normalsize
where $k\in \mathcal{I}_I$. After this transformation, the integral on the right side of Eq.(\ref{eq:CG_force_field}) can be integrated by parts and the following partial derivative will emerge,
$$\frac{\partial}{\partial \bm{r}_{k}} \left[ \text{exp}\left(-\frac{u(\bm{r}^n)}{k_\text{B}T}\right) \prod_{J\neq I}\delta\big(\bm{M}_{\bm{R}_J}(\bm{r}^n)\allowbreak-\bm{R}_J\big) \right]~.$$
This partial derivative can be greatly simplified if we focus on atoms that are only mapped to the $I$-th coarse-grained bead but not other beads. That is, the index $k$ is limited to all $k\in \mathcal{I}_I$ but $k\notin \mathcal{I}_J$ for $J \neq I$. For such indices, a nonzero factor ${d_{Ik}}$ can be introduced such that
\begin{align}\label{eq:d_factor}
    \sum_{k\in \mathcal{S}_I}d_{Ik}=1 
    \text{\hspace{5 pt} for all~} I~,
\end{align}
where $\mathcal{S}_I$ is the set of indices for the atoms that only belong to the $I$-th coarse-grained bead in the coarse-graining mapping. The combination of Eq.~(\ref{eq:partial_derivative_transform}) and Eq.~(\ref{eq:d_factor}) yields the following identity,
\footnotesize
\begin{align}\label{eq:partial_derivative_identity}
    \frac{\partial}{\partial \bm{R}_I}\delta \left(\sum_{i\in \mathcal{I}_I} c_{Ii}\bm{r}_i-\bm{R}_I \right)   = - \sum_{k\in\mathcal{S}_I}\frac{d_{Ik}}{c_{Ik}}\frac{\partial}{\partial\bm{r}_k}\delta \left(\sum_{i\in\mathcal{I}_I}c_{Ii}\bm{r}_i-\bm{R}_I \right)~.
\end{align}
\normalsize

Using Eq.~(\ref{eq:partial_derivative_identity}), Eq.~(\ref{eq:CG_force_field}) can be rewritten as
\small
\begin{align}\label{eq:CG_force_field_2}
    \bm{F}_I(\bm{R}^N) = &-\sum_{k\in\mathcal{S}_I}\frac{d_{Ik}}{c_{Ik}}\frac{k_{B}T}{z(\bm{R}^N)}\int d\bm{r}_{\overline{k}}^{n-1}
    \prod_{J\neq I}\delta(\bm{M}_{\bm{R}_J}(\bm{r}^n)-\bm{R}_J) \nonumber\\
    &  \int d\bm{r}_k \text{exp}\left(\frac{-u(\bm{r}^n)}{k_\text{B}T}\right) \frac{\partial}{\partial\bm{r}_k}
    \delta \left(\sum_{i\in\mathcal{I}_I}c_{Ii}\bm{r}_i-\bm{R}_I \right)~,
\end{align}
\normalsize
where $d\bm{r}_{\overline{k}}^{n-1} \equiv \prod_{j=1, j\neq k}^n d\bm{r}_j$. An integration by parts yields
\small
\begin{align} \label{eq:CG_force_ensemble}
    \bm{F}_I(\bm{R}^N) = & \frac{1}{z(\bm{R}^N)}\int d \bm{r}^n \text{exp} \left( \frac{-u(\bm{r}^n)}{k_\text{B}T} \right)
    \delta \left(\bm{M}_{\bm{R}^N}(\bm{r}^n)-\bm{R}^N \right)\nonumber\\
    & \sum_{k\in\mathcal{S}_I}\left(-\frac{d_{Ik}}{c_{Ik}}\right) \frac{\partial u(\bm{r}^n)}{\partial\bm{r}_k} \nonumber\\
    =& \langle \sum_{k\in\mathcal{S}_I} \frac{d_{Ik}}{c_{Ik}} \bm{f}_k(\bm{r}^n) \rangle_{\bm{R}^N}~,
\end{align}
\normalsize
where $\bm{f}_k(\bm{r}^n) \equiv  - \frac{\partial u(\bm{r}^n)}{\partial\bm{r}_k}$ represents the force on the $k$-th atom from all other atoms in the system and $\langle \cdots \rangle_{\bm{R}^N}$ indicates an ensemble average performed under the coarse-graining mapping encoded in Eq.~(\ref{eq:CG_mapping_coord}).

Equation~(\ref{eq:CG_force_ensemble}) is the theoretical foundation of computing the coarse-grained force field. In the mapping adopted here, all atoms are separated into nonoverlapping groups, each of which is mapped to one and only one coarse-grained bead. Furthermore, the coarse-grained bead is placed at the center of mass of the atomic group that it represents. In this case, $c_{Ik} =\frac{m_k}{\sum_{j\in \mathcal{S}_I} m_j}$ and if $k\in \mathcal{S}_I$, then $k\notin \mathcal{S}_J$ for all $J \neq I$. A natural choice is $d_{Ik} = c_{Ik}$, which indicates that the force on the $I$-th coarse-grained bead is the ensemble average of the forces on atoms with indices in $\mathcal{S}_I$ from all other atoms in the system, subjected to the mapping from atoms to beads. Noid \textit{et al.} further proved that this scheme also guarantees that the momentum part of the phase-space probability distribution function in the coarse-grained model matches that in the atomistic model.\cite{Noid2008a}

To summarize, the coarse-grained model and the atomistic model have consistent probability distribution functions of thermodynamic states in the phase space as long as
\begin{itemize}
    \item One group of atoms is mapped to one coarse-grained bead.
    \item No atom is shared by more than one group.
    \item A coarse-grained bead is placed at the center of mass of the atomic group that the bead represents.
    \item The force on a coarse-grained bead is the ensemble average of the forces exerted on all atoms in the group, which the bead represents, by all other atoms in the atomistic system.
\end{itemize}
The coarse-graining method on the basis of these constraints is called force-matching coarse-graining.\cite{Ercolessi1994, Izvekov2005JCP, Izvekov2005JPCB, Noid2008a, Noid2008b, Wang2009}

\subsection{Pairwise Nonbonded Coarse-Grained Force Field}

The force-matching theory of coarse-graining discussed in the previous section is valid for any molecular system. However, the resulting coarse-grained force field is a many-body force field, which is very difficult to compute in general.\cite{Noid2008a, Noid2008b, Wang2009} Furthermore, the coarse-grained model is state-dependent as the coarse-grained force field is computed for a given thermodynamic state, as indicated by Eq.~(\ref{eq:CG_force_ensemble}). Therefore, the coarse-grained model developed using the general theory is usually not transferable, though it is proved to be self-consistent and rigorous from the perspective of thermodynamics.\cite{Izvekov2005JCP, Izvekov2005JPCB, Noid2008a, Noid2008b}

In order to improve the transferability of the resulting coarse-grained models, a mean-field approximation to the force-matching coarse-graining approach was proposed by Wang and collaborators.\cite{Wang2009} Their effective-force coarse-graining (EF-CG) method results in a pairwise coarse-grained force field. In EF-CG, an atomic system is divided into groups of atoms, with each group mapped to a coarse-grained bead. The interaction between a pair of beads is computed from the average force between the corresponding atomic groups. When this calculation is performed in a condensed phase including all other atoms, the many-body nature of the coarse-grained force field is naturally captured.\cite{Wang2009} However, this type of implementation of EF-CG requires MD simulations of the entire atomistic system and nontrivial constraints on the pair of atomic groups being parameterized.

In this paper, we adapt the EF-CG method\cite{Wang2009} and propose an efficient gas-phase sampling scheme to compute the pairwise PMFs between atomic groups, in order to develop a transferable and expandable coarse-grained model for a branched polyetherimide. In particular, the former feature indicates that the model can be used to simulate the polyetherimide under different temperatures and the latter means that when a new polyetherimide containing one or several new functional groups is dealt with, the same number of new coarse-grained beads will be added to the existing coarse-grained model. The only parameterization that is needed to update the coarse-grained model is to compute the interactions between the new beads and all the existing beads. In this sense, the coarse-grained force field is expanded with the newly added beads. This library-like approach is obviously appealing as it avoids the need to construct a new coarse-grained model every time when the polymer of interest is updated. The general theoretical frameworks of deriving libraries of transferable coarse-grained potentials were discussed by Mullinax and Noid previously, who proposed an extended ensemble approach and combined it with the force-matching method to derive transferable coarse-grained models for molecular liquids\cite{Mullinax2009} and proteins.\cite{Mullinax2010} Recently Sanyal \textit{et al.} has adapted this approach to the relative entropy framework to develop transferable coarse-grained protein models.\cite{Sanyal2019} For polymeric materials, more systematic work is still needed to establish the applicability of a bottom-up, library-like approach to construct transferable and expandable coarse-grained models.\cite{Noid2013}

The simplest system is one that can be divided into two groups of atoms and can be coarse-grained into two beads. The force on one bead is then the sum of the forces on all atoms in the group this bead represents by all atoms in the other group. If the center of mass of each group is chosen as the location of the corresponding coarse-grained bead, then the coarse-grained model with a pairwise interaction is fully consistent with the multiscale coarse-graining framework proposed by Voth, Andersen, Noid \textit{et al.}.\cite{Noid2008a} This is easy to understand as in such a simplified system, there is no many-body effect since there are only two coarse-grained beads anyway. In a more general case, there are of course more than two beads in the coarse-grained model. Then parameterizing the coarse-grained force field in a pairwise fashion assumes the force field is additive and automatically neglects its intrinsic many-body nature as required by the thermodynamic consistency between the atomistic and coarse-grained models. However, the many-body correlations are still important and have to be incorporated into the final coarse-grained force field,\cite{Rudzinski2012, Scherer2018} albeit implemented in a pairwise manner. In this paper we show that a correction term can be added to the pairwise PMFs derived with gas-phase MD calculations to not only make the model transferable but also to some extent compensate for the error caused by the neglected many-body effects.

\subsection{Fixing the Center of Mass of a Group of Atoms}

When computing the force between two groups of atoms, one technical key is to fix the center of mass of each group, which will then allow the separation between the two centers of mass to be used as a coarse-grained coordinate. Fritz \textit{et al.} used the LINCS constraint algorithm to fix the center-of-mass separation of two groups and compute their mutual interaction.\cite{Fritz2009} Here, we adopt a different approach implemented in LAMMPS. In the starting configuration, the velocity of the center of mass of each atomic group is set to zero (i.e., the total momentum of the group is set to zero). When a group of atoms interact with atoms from other groups, all atoms first move according to the Newtonian equation of motion. After one time step in MD simulations, the atomic coordinates are updated and the new location of the center of mass of each group is computed. The displacement vector, $\vec{d}_c$, of a group's center of mass from its original location to the new one is determined as well as its velocity, $\vec{v}_c$. Then, all atoms in that group are displaced by $-\vec{d}_c$ such that the center of mass of the group is shifted back to its original location. At the same time, $\vec{v}_c$ is subtracted from the velocity of each atom in the group to ensure that the group's center-of-mass velocity is restored to zero. In the Supporting Information, we prove that this ``recentering'' approach is equivalent to the method where a constraint force is added to each atom in a group such that the group's center of mass does not move (i.e., the total force on the group, including the constraint forces applied to all the atoms in the group, is always zero) and the group does not exhibit any artificial rotation (i.e., the net torque from the constraint forces on the group is zero). Herein we employ the ``recentering'' approach to constrain the center of mass of an atomic group.

\subsection{Sampling Configurations at a Fixed Center-of-Mass Distance: Test}

\begin{figure}[htb]
    \centering
    \includegraphics[width=0.45\textwidth]{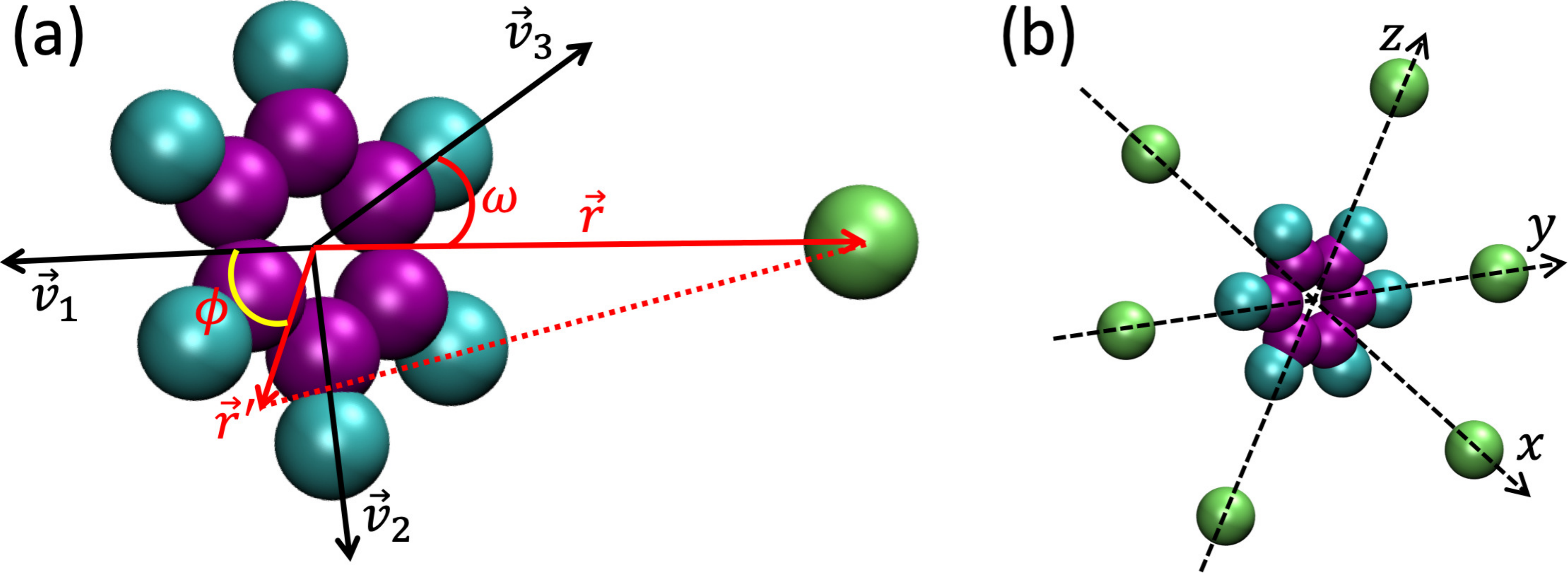}
    \caption{(a) Snapshot of a model system consisting of one benzene molecule and one oxygen atom. (b) Snapshot of a model system consisting of one benzene molecule at the center and six surrounding oxygen atoms in the $+x$, $-x$, $+y$, $-y$, $+z$, and $-z$ directions.}
    \label{fig:benzene_oxygen}
\end{figure}

To demonstrate the method of using PMF calculations to approximate the ensemble average in Eq.~(\ref{eq:CG_force_ensemble}), we utilize a model system that consists of one benzene molecule and one oxygen atom, as shown in Fig.~\ref{fig:benzene_oxygen}(a). A configuration of this model system can be approximately characterized by three parameters, the magnitude of the vector $\vec{r}$ pointing from the benzene's center of mass to the oxygen and two angles that describe the orientation of $\vec{r}$ relative to the benzene molecule. A Cartesian coordinate system can be set up using the benzene's center of mass as its origin and three orthonormal vectors, $\vec{v}_1$, $\vec{v}_2$, and $\vec{v}_3$, as the axes. Among these, $\vec{v}_1$ and $\vec{v}_2$ define the plane in which the benzene molecule lies in and $\vec{v}_3$ is normal to this plane. At a finite temperature, the six carbon atoms and six hydrogen atoms in the benzene molecule actually have a three-dimensional conformation. However, we can always define a plane that captures the planar nature of the benzene molecule. For example, a plane that minimizes the sum of square distances or has zero average distance for all the carbon atoms from this plane can be used. The location of the oxygen atom in this coordinate system is therefore given by the vector $\vec{r}$, i.e., its length $r \equiv |\vec{r}|$ as well as two angles: the polar angle $\omega$ and the azimuthal angle $\phi$, as shown in Fig.~\ref{fig:benzene_oxygen}(a).

We built all atomistic systems using MAPs.\cite{maps} The bond increment method was used to set the partial charges of atoms.\cite{halgren1996merck} The PCFF force field was adopted for all-atom MD simulations,\cite{sun1994} which was used in several previous studies of polyetherimides.\cite{YoungHinkley1999,Lim2003,Lim2007} The equation of motion was integrated using the velocity-Verlet algorithm with the time step set as 1 fs. The cutoff of both van der Waals and Coulomb interactions was set as 12 \AA. The particle-particle particle-mesh (PPPM) method was used to calculate the long-range part of Coulomb interactions. When needed, a Langevin thermostat and a Berendsen barostat were used to control temperature and pressure, respectively.

The ``recentering'' method of fixing a center of mass discussed previously is applied to the benzene-oxygen system and used to fix $r$ at any chosen value that is physically allowed. At a given $r$, we use all-atom MD simulations to sample various configurations parameterized by $\omega$ and $\phi$ and compute the probability density, $P(\omega, \phi)$, of a given configuration. Considering the rotational symmetry of a benzene molecule around $\vec{v}_3$, we investigate the probability density integrated over $\phi$, i.e., $P(\omega) \equiv\frac{1}{2\pi} \int_0^{2\pi}P(\omega, \phi) d\phi$.

We compute $P(\omega)$ in five different ways. First, it is directly estimated using the trajectory (i.e., a series of configurations) generated by a MD simulation at a constant temperature $T$. Namely, $P(\omega)d\omega = \frac{\#[\omega,\omega+d \omega]}{\text{total number of states}}$, where $\#[\omega,\omega+d \omega]$ is the number of states with the polar angle in $[\omega, \omega + d\omega]$. Second, the energy of each configuration, $\epsilon(\omega, \phi)$, from the MD simulation is used to compute $P(\omega)$ through the canonical distribution, $P(\omega) = \frac{\sum_\phi e^{-\beta \epsilon(\phi,\omega)}\sin \omega}{\sum_\omega \sum_\phi e^{-\beta\epsilon(\phi,\omega)}\sin\omega}$, where $\beta = 1/(k_\text{B}T)$ and $k_\text{B}$ is the Boltzmann constant. Third, we compute $\epsilon(\phi,\omega)$ for a series of static configurations of the single benzene-oxygen pair. Noting the symmetry of the system, the polar angle $\omega$ is varied from 0 to $\pi/2$ in steps of $\pi/90$. For each $\omega$, the azimuthal angle $\phi$ is varied from 0 to $\pi$ in increments of $\pi/90$. For each configuration, the benzene molecule is in its ground-state conformation (i.e., a planar hexagonal configuration) and the interaction energy between the benzene molecule and the oxygen atom, $\epsilon(\omega, \phi)$, is computed. The canonical distribution is then used to determine $P(\omega)$. Additionally, $P(\omega)$ is determined with the trajectory or energy generated from a MD simulation of a model system that consists of one benzene molecule at the origin of a Cartesian coordinate system and six oxygen atoms surrounding it at the same separation but in the positive and negative directions of the three axes, as shown in Fig.~\ref{fig:benzene_oxygen}(b). This system is designed to speed up the sampling of various configurations, in particular, rare configurations that may have significant contributions to the ensemble average of mutual forces when $r$ is small. In this case, extremely long MD simulations are needed if a single pair of benzene and oxygen is used. Later in this paper we show that this strategy of simultaneously using six pairs is practically useful when PMF calculations are performed for atomic groups whose conformations deviate significantly from a sphere, such as aromatic rings.

\begin{figure}[htb]
    \centering
    \includegraphics[width=0.45\textwidth]{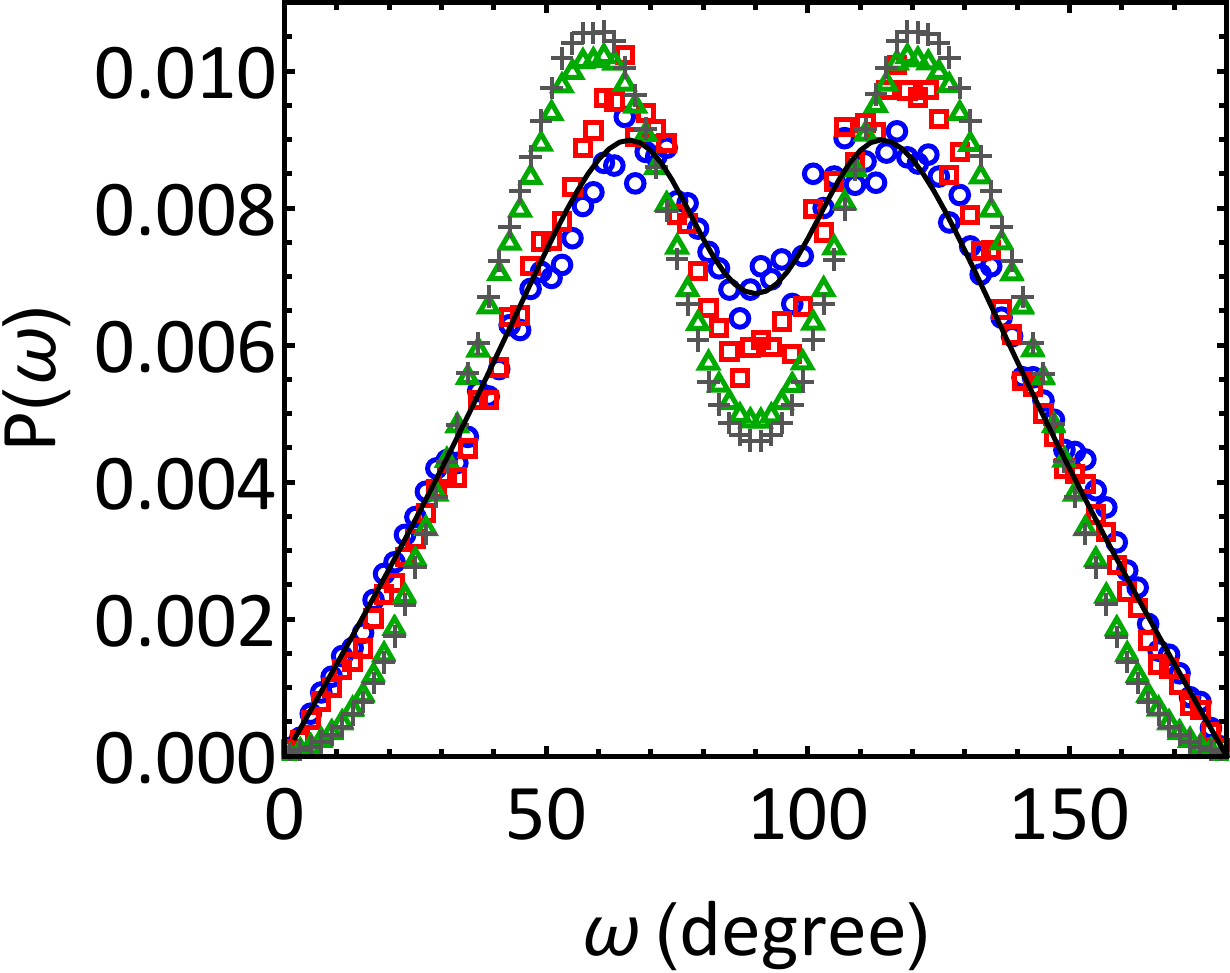}
    \caption{Comparison of the probability density, $P(\omega)$, calculated in five different ways for a benzene-oxygen system at $r=5$ \AA. The data are based on MD trajectory ($\bigcirc$) and energy ($\square$) of the single benzene-oxygen pair at $T=300$ K, MD trajectory ($\triangle$) and energy (+) of the one benzene/six oxygen system at $T=300$ K, and static configuration energy (solid line) of the single benzene-oxygen pair (effectively at $T=0$ K).}
    \label{fig:prob_den_benzene_oxygen}
\end{figure}

The results of $P(\omega)$ at $r=5$ \AA~are shown in Fig.~\ref{fig:prob_den_benzene_oxygen}. For both the single benzene-oxygen pair and the one benzene/six oxygen system, the results based on the MD trajectory (circles and triangles in Fig.~\ref{fig:prob_den_benzene_oxygen}) and the corresponding energy (squares and pluses in Fig.~\ref{fig:prob_den_benzene_oxygen}) are close but differ slightly around $\omega = 60^{\circ}$, $90^{\circ}$, and $120^{\circ}$. This difference is likely due to the limited number of configurations sampled in the MD simulations as it can be noted that the difference between the two (triangles and pluses in Fig.~\ref{fig:prob_den_benzene_oxygen}) is much smaller for the one benzene/six oxygen system in Fig.~\ref{fig:benzene_oxygen}(b), which can sample more configurations in the same number of MD time steps. The results computed using the energy of uniformly-scanned static configurations (solid line in Fig.~\ref{fig:prob_den_benzene_oxygen}), on the other hand, show a very good agreement with those based on the MD trajectory of the single benzene-oxygen pair. This agreement may indicate that for the single benzene-oxygen pair at 300 K, the only unconstrained degrees of freedom of the benzene ring, after the rotation around its fixed center of mass is subtracted, are very weak bond vibrations and the energy $\epsilon(\omega, \phi)$ at 300 K is therefore close to the corresponding value at 0 K.

The results of the system with six benzene-oxygen pairs are in reasonable agreement with those from one pair, though the former seem to slightly overestimate (oversample) configurations with $30^{\circ} \lesssim \omega \lesssim 60^{\circ}$ and $120^{\circ} \lesssim \omega \lesssim 150^{\circ}$, while underestimate (undersample) configurations in the other ranges of $\omega$. Although the discrepancy is noted, we will utilize the six-pair setup in Fig.~\ref{fig:benzene_oxygen}(b) to speed up the calculation of pairwise nonbonded interactions between coarse-grained beads when developing the coarse-grained model of the branched polyetherimide below. The error introduced by this choice will be balanced out when a correction term is included in the coarse-grained force field, as discussed later.

The PMF sampling scheme as shown in Fig.~\ref{fig:benzene_oxygen}(a) is reminiscent of the conditional reversible work (CRW) method developed by van der Vegt and coworkers.\cite{Fritz2009, brini2011conditional, deichmann2017conditional, Deichmann2019} In the CRW approach, the PMF between atomic groups $I$ and $J$ is computed with two oligomers. One oligomer contains $I$ while the other contains $J$. At a fixed center-of-mass distance between $I$ and $J$, the force between the oligomers is computed with the direct $I$-$J$ interaction either turned on or off. The difference between the two is the effective interacting force between $I$ and $J$ in a many-body environment and its integration over the center-of-mass separation yields the sought PMF. The force sampling can be performed in a vacuum either with MD \cite{Fritz2009, brini2011conditional, deichmann2017conditional} or a Monte Carlo scheme.\cite{Deichmann2019} PMFs from the CRW and EF-CG approaches are very close as the two share a similar principle,\cite{Deichmann2019comparison} though in EF-CG the trajectory sampling is usually performed in a condensed phase. We tested CRW for the branched polyetherimide targeted in this paper but the results were unsatisfactory, potentially because of the polymer's branched topology. As a result, we turn to the six-pair sampling scheme in Fig.~\ref{fig:benzene_oxygen}(b) to compute PMFs with gas-phase MD simulations, which to some extent is an intermediate between CRW (i.e, a pair of oligomers in a vacuum)\cite{brini2011conditional} and EF-CG (i.e, multiple pairs of atomic groups in a condensed phase).\cite{Wang2009}

\section{Development of a Coarse-Grained Model of a Branched Polyetherimide} \label{sec:CG_model_development}

\begin{figure*}[htb]
  \centering
  \includegraphics[width=0.7\textwidth]{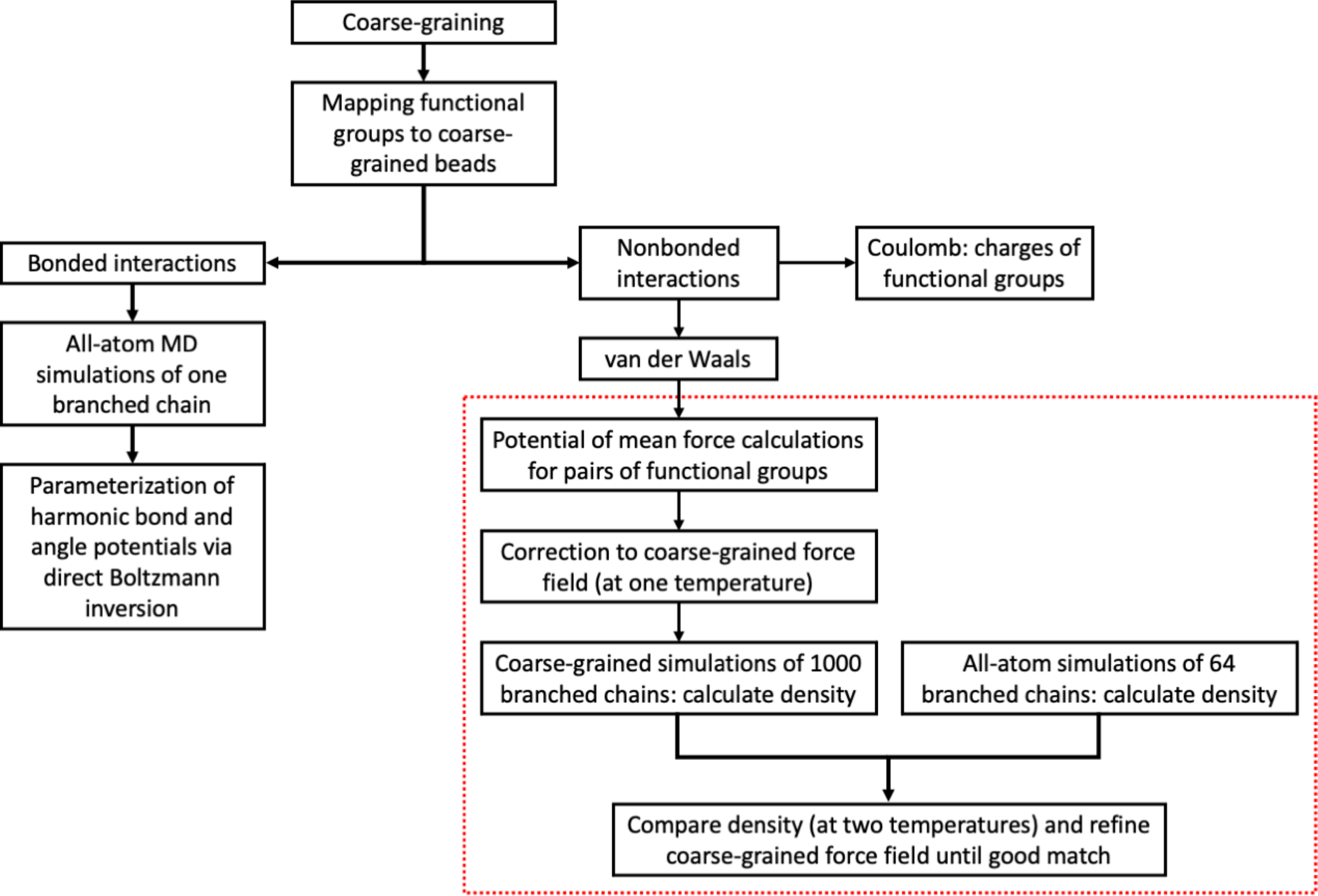}
  \caption{The coarse-graining flowchart.}
  \label{fig:CG_scheme}
\end{figure*}

We employ the methodology discussed in the previous section to develop a coarse-grained model of a branched polyetherimide derived from Ultem and use this model to compute its mechanical, structural, and rheological properties. The protocol of developing such a model is outlined in Fig.~\ref{fig:CG_scheme}, which mainly includes three steps: the grouping of atoms into coarse-grained beads, the parameterization of bonded (i.e., bond, angle, and dihedral) interactions, and the parameterization of nonbonded interactions between the coarse-grained beads. In the following sections we describe each step in detail.

\subsection{Mapping Groups of Atoms into Coarse-Grained Beads} \label{sec:CG_mapping}

\begin{figure}[htb]
  \centering
  \includegraphics[width=0.45\textwidth]{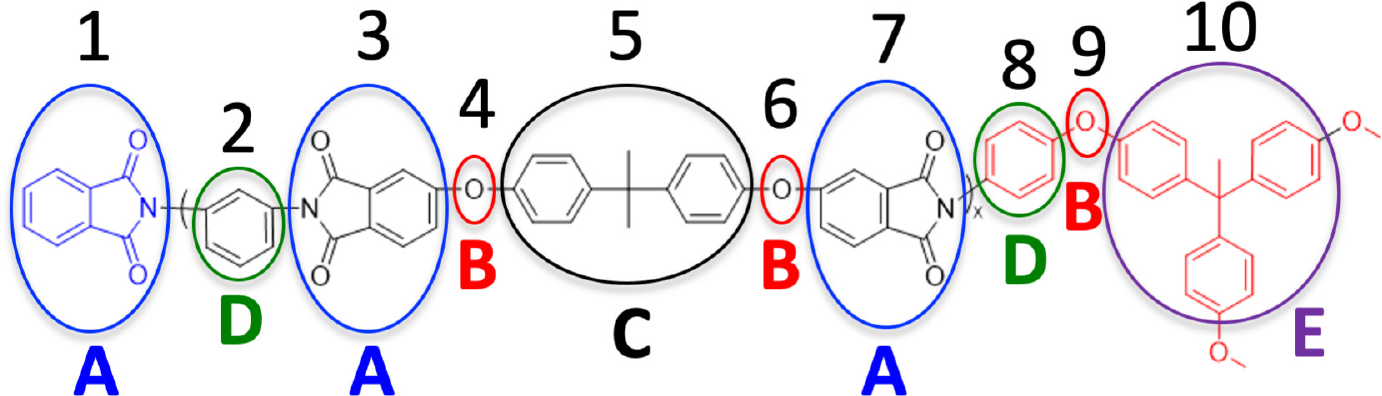}
  \caption{Mapping atomic groups into five types of coarse-grained beads. For clarity, only one short branch connected to TAPE and terminated with PA is shown.}
  \label{fig:CG_map}
\end{figure}

The branched polyetherimide studied here is polymerized from two backbone monomers, BPADA and MPD, a chain terminator, PA, and a tri-functional branching agent, TAPE. One short branch, terminated on one end with PA and connected to TAPE on its other end, is shown in Fig.~\ref{fig:CG_map}. After several attempts, we settle with using 5 types of coarse-grained beads for such a system: type-A beads for the phthalimide groups, type-B beads for the oxygen atoms in the flexible ether linkages, type-C beads for the bisphenol A groups, type-D beads for the aromatic rings in the phenylenediamine groups, and type-E beads for the core parts of TAPEs. The groupings are shown schematically in Fig.~\ref{fig:CG_map}. In total there are 10 atomic groups in this short branch. Group 1 is chemically almost the same as groups 3 and 7 except that there is one extra hydrogen atom in group 1 since it is at the end of the branch. The mass and charge of group 1 are therefore slightly different from those of groups 3 and 7. However, we map all three groups to type-A beads (with slightly different masses and charges) to simplify the non-Coulombic part of their nonbonded interactions with other coarse-grained beads. The error of this treatment is partially compensated for by a correction to the coarse-grained force field introduced later. Groups 2 and 8 are chemically almost identical except for the location of one hydrogen atom. As a result, they have the same mass but their charges are somewhat different. For simplicity, we map these groups into type-D beads with different charges. The charges and masses of all atomic groups are summarized in Table~\ref{tb:cg_charge}.

\begin{table}[htb]
\centering
\caption{Charges and masses of atomic groups (AG) defined in Fig.~\ref{fig:CG_map} and their mapping to coarse-grained (CG) beads.}
\label{tb:cg_charge}
\begin{tabular}{|l|l|l|l|}
\hline
CG Bead & AG Index & Charge ($e$) & Mass ($10^{-25}$ g)\\ \hline
\multirow{2}{*}{type-A} & 1 & 0.05 & 2426.46 \\ \cline{2-4}
                        & 3,~7 & 0.0765 & 2409.72\\ \hline
type-B & 4,~6,~9 & -0.053 & 265.67\\ \hline
type-C & 5 & 0.053 & 3226.05\\ \hline
\multirow{2}{*}{type-D} & 2 & -0.1 & \multirow{2}{*}{1263.64} \\ \cline{2-3}
                        & 8 & -0.0235 & \\ \hline
type-E & 10 & 0.0795 & 4240.02\\ \hline
\end{tabular}
\end{table}

We adopt the chemistry-informed mapping scheme in Fig.~\ref{fig:CG_map} because it leads to unimodal distributions for all the bonds and angles between the coarse-grained beads and therefore a Gaussian approximation can be used to derive the stiffness constants of bonds and angles, as discussed below. Other mapping schemes, including those more coarsened, usually cause the angle potential to have two or more local minima, which is not desirable for the parameterization of the bond and angle interactions between the coarse-grained beads. We also notice that it is necessary to keep the oxygen atoms in the ether linkages as a distinct type of coarse-grained beads. This observation is consistent with the understanding that the inclusion of ether groups in polyetherimides is responsible for their enhanced chain flexibility and improved melt processability compared with other polyimides.\cite{Urakawa2014}

\subsection{Parameterization of Coarse-Grained Bonded Interactions} \label{sec:bonded}

The second step of coarse-graining is to parameterize the bond and angle interactions in the coarse-grained chain using results from all-atom MD simulations with a single branched polyetherimide chain. To keep the coarse-grained model simple, dihedral interactions are not included. For two neighboring groups of atoms, each mapped to a coarse-grained bead, we compute their center-of-mass distance and examine its probability distribution. For three consecutive groups of atoms, we compute the angle formed by the corresponding centers of mass and examine the probability distribution of the angles. All probability distributions based on the grouping scheme in Fig.~\ref{fig:CG_map} are well approximated by Gaussian distributions, indicating that the bonds and angles can be described by a harmonic potential,
\begin{align}\label{eq:harmonic_pot}
U(x)=\frac{1}{2}k_x(x-x_0)^2~,
\end{align}
and the corresponding probability density is
\begin{align}\label{eq:direct_Boltzmann_inversion}
p(x)=\sqrt{\frac{k_x}{2\pi k_\text{B} T}}e^{-\frac{U(x)}{k_\text{B} T}}~,
\end{align}
where $x$ is the length $r$ for a bond or the angle $\theta$ for an angle, $x_0$ the equilibrium bond length or angle, $k_x$ the corresponding stiffness, and $U(x)$ the harmonic bond or angle potential. The method of using Eq.~(\ref{eq:direct_Boltzmann_inversion}) to derive $U(x)$ for bonded interactions from $p(x)$ is called direct Boltzmann inversion.\cite{Fritz2009} When $p(r)$ and $p(\theta)$ are evaluated, the corresponding Jacobian factors $r^2$ and $\text{sin}\theta$ need to be included.\cite{Tschop1998, Fritz2009}

\begin{figure}[htb]
  \centering
  \includegraphics[width= 0.4\textwidth]{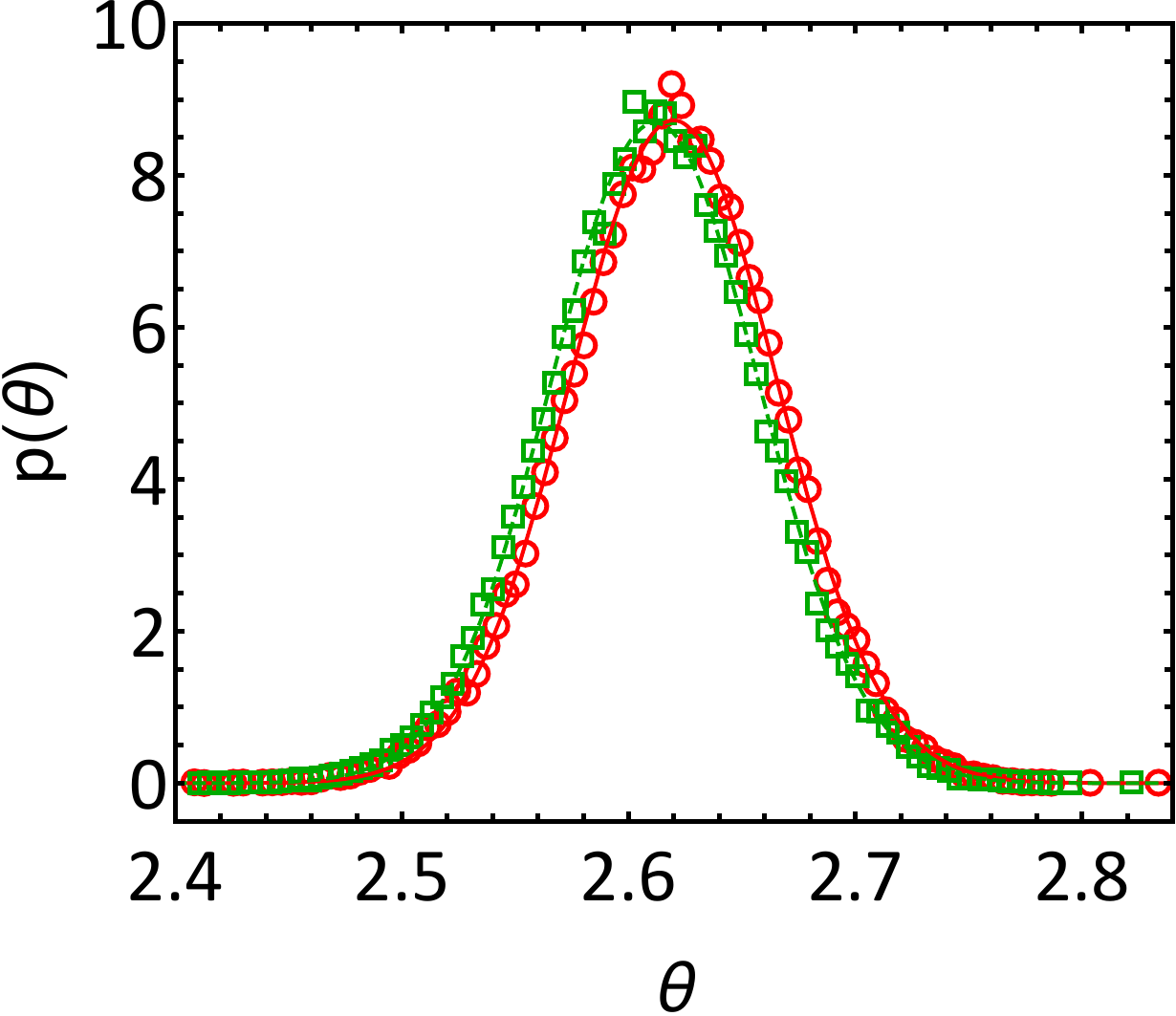}
  \caption{The probability distribution of the 2-3-4 angle at $T=450$ K: all-atom MD data (circles) and a Gaussian fit (solid line) with $k_{\theta}$=427 kcal/mol/rad$^2$ and $\theta_0$=2.62 rad; the probability distribution of the 6-7-8 angle at the same $T$: all-atom MD data (squares) and a Gaussian fit (dashed line) with $k_{\theta}$=425 kcal/mol/rad$^2$ and $\theta_0$=2.61 rad.}
  \label{fig:angle_dist}
\end{figure}

Figure \ref{fig:angle_dist} shows the probability distributions of the 2-3-4 angle and the 6-7-8 angle (see Fig.~\ref{fig:CG_map} for the definition of bead types and atomic group indices), which are both D-A-B type angles. The probability densities are well fit by Gaussian distributions. The fitting allows us to extract a spring constant and an equilibrium angle for the harmonic angle potential. The spring constants and equilibrium values for the same type angles are very close, as shown in the example in Fig.~\ref{fig:angle_dist}. Similar results are obtained for all bonds and angles. The spring constants, equilibrium bond lengths, and equilibrium angles for all bonds and angles are summarized in Table~\ref{tb:bond_parameter} and Table~\ref{tb:angle_parameter}. The results validate the usage of harmonic potentials for the bond and angle interactions in the coarse-grained force field. For the same type bonds and angles appearing more than once even in one branched chain, they all have similar stiffnesses and equilibrium values. This finding justifies the simplification of mapping 10 atomic groups into 5 types of coarse-grained beads. We further confirm that the parameters of the coarse-grained bond and angle potentials are insensitive to temperature for the range of temperatures of interest here. The last 2 columns of Table~\ref{tb:bond_parameter} and Table~\ref{tb:angle_parameter} contain the final parameters of the coarse-grained bond and angle potentials.

\begin{table}[htb]
\centering
\caption{Coarse-grained bond parameters.}
\label{tb:bond_parameter}
\begin{tabular}{|c|c|c|c|c|c|}
\hline
Bond& Bonds & $k_r$ & $r_0$ & $\overline{k}_r$ &$\overline{r}_0$\\
 Type &  & (kcal/mol/\AA$^2$) & (\AA)  & (kcal/mol/\AA$^2$) & (\AA) \\
\hline
\multirow{3}{*}{A-D} & 1-2 & 267 & 4.86 & \multirow{3}{*}{270} &\multirow{3}{*}{4.88}\\ \cline{2-4}
                              & 2-3 & 255 & 4.87&  &\\ \cline{2-4}
                            & 7-8 & 286 & 4.91& &\\
\hline
\multirow{2}{*}{A-B} & 3-4 & 381 & 4.09 &\multirow{2}{*}{391}&\multirow{2}{*}{4.08}\\ \cline{2-4}
                                & 6-7 & 400 & 4.07& & \\
\hline
\multirow{2}{*}{B-C} & 4-5 & 88 & 4.99 &\multirow{2}{*}{86}&\multirow{2}{*}{5.02}\\  \cline{2-4}
                                 & 5-6 & 83 & 5.04 & &\\
\hline
B-D & 8-9 & 505 & 2.80& 505& 2.80 \\
\hline
B-E & 9-10 &80 & 5.58& 80& 5.58 \\
\hline
\end{tabular}
\end{table}

\begin{table}[htb]
\centering
\caption{Coarse-grained angle parameters.}
\label{tb:angle_parameter}
\begin{tabular}{|c|c|c|c|c|c|}
\hline
Angle& Angles & $k_\theta$ & $\theta_0$  & $\overline{k}_\theta$ & $\overline{\theta}_0$\\
Type &  & (kcal/mol/rad$^2$) & (rad)  & (kcal/mol/rad$^2$) & (rad) \\
\hline
A-D-A & 1-2-3 & 254 & 2.10 &254&2.10\\
\hline
\multirow{2}{*}{D-A-B}  & 2-3-4 & 427 & 2.62 & \multirow{2}{*}{426} &\multirow{2}{*}{2.62} \\ \cline{2-4}
                                     & 8-7-6 & 425 & 2.61& & \\
\hline
\multirow{2}{*}{A-B-C} & 3-4-5 & 74 & 2.17&\multirow{2}{*}{59}&\multirow{2}{*}{2.12}   \\ \cline{2-4}
                                    & 7-6-5 & 44 & 2.07& & \\
\hline
B-C-B & 4-5-6 & 124 & 1.70 &124&1.70\\
\hline
A-D-B & 7-8-9 & 108 & 2.23 &108&2.23\\
\hline
D-B-E & 8-9-10 & 507 & 3.11 &507&3.11\\
\hline
B-E-B & 9-10-9 & 79 & 2.20 &79&2.20\\
\hline
\end{tabular}
\end{table}

\subsection{Parameterization of Coarse-Grained Nonbonded Interactions}

\subsubsection{Potential of Mean Force Calculations}

The next step is to parameterize the nonbonded interactions between the coarse-grained beads. In each PMF calculation, the center of mass of each atomic group is fixed but the atoms in the group are still mobile, causing the group to rotate and vibrate around its center of mass. A Langevin thermostat is used to keep the system at a target temperature. By sufficiently sampling the relative configurations and orientations of the two groups with their centers of mass separated at a given distance, we calculate the average force between them as a function of the separation. The results show that the average force is along the vector connecting the two centers of mass. Therefore the separation between the centers of mass can be used as a coarse-graining coordinate. Integrating the average force over the center-of-mass separation, we obtain the coarse-grained potential for each pair of beads.

Since atoms carry charges in an all-atom model, we split the coarse-grained potential into two parts: the Coulomb component and the van der Waals component. Coulomb interactions are included in the coarse-grained force field by using the total charge of a group as the charge of the corresponding coarse-grained bead. The charges of all atomic groups for the branched polyetherimide can be found in Table~\ref{tb:cg_charge}. The Coulomb force is subtracted from the mean force between the two groups and the remaining part is designated as the van der Waals component, which is still along the vector connecting the centers of mass. We call the integration of this component over the center-of-mass separation the nonbonded, van der Waals PMF. Its mathematical expression is
\begin{equation} \label{eq:PMF_integral}
U(r) = -\int_{r_m}^r \langle f_c \rangle_s ds,
\end{equation}
where $r_m$ is a large center-of-mass separation at which $U(r_m) \simeq 0$,  and $\langle f_c \rangle_s$ is the force between the two groups at separation $s$ with the Coulomb force between them subtracted. Hereafter a PMF always refers to the van der Waals component of the nonbonded interaction between two atomic groups. Eq.~(\ref{eq:PMF_integral}) has been used in CRW\cite{brini2011conditional} and EF-CG.\cite{Deichmann2019comparison} Other researchers have also used Eq.~(\ref{eq:PMF_integral}) to compute the mean potentials between ions,\cite{Ciccotti1989, Guardia1991} molecules,\cite{Pranami2009} and nanoparticles.\cite{Schapotschnikow2009, Bauer2015, Munao2018}

In our approach, the force $\langle f_c \rangle_s$ in Eq.~(\ref{eq:PMF_integral}) is computed for a pair of atomic groups in a vacuum. This is different from the sampling scheme adopted in CRW\cite{brini2011conditional} and EF-CG.\cite{Deichmann2019comparison}, where the constraint force needed to maintain two atomic groups at a given center-of-mass separation in the presence of other atomic groups is computed and used as $-\langle f_c \rangle_s$ in Eq.~(\ref{eq:PMF_integral}). Our gas-phase sampling approach to determine PMFs thus assumes that the coarse-grained force field can be decomposed into pair-additive PMFs, which neglects the many-body correlations present in a condensed phase. However, the coarse-grained force field parameterized in this way has the advantage of being easily expandable. Later we will show that a correction can be introduced to the PMFs to account for the error caused by the ignored many-body correlations and to enable the coarse-grained model to be temperature transferable.

\begin{figure}[htb]
  \centering
  \includegraphics[width=0.4\textwidth]{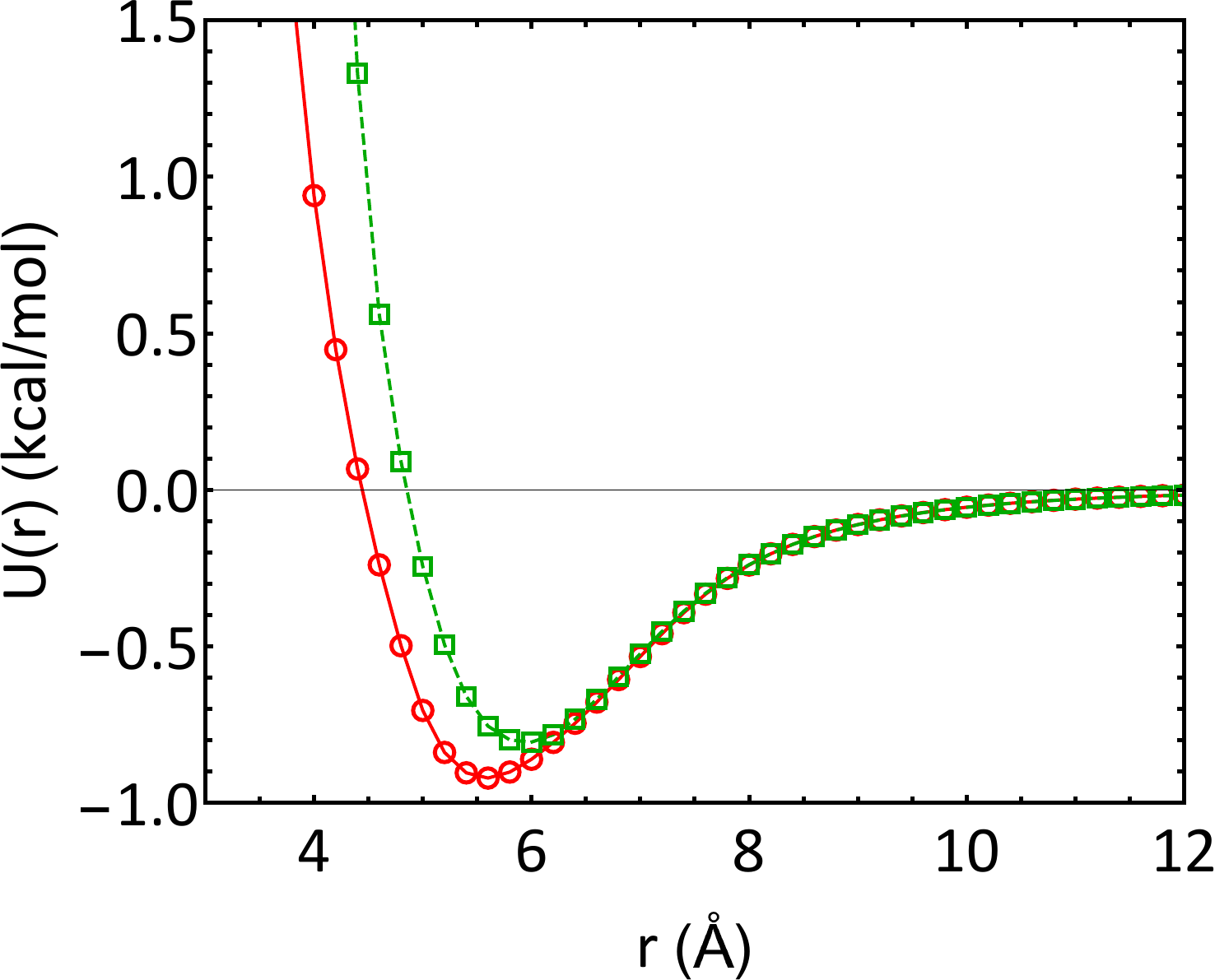}
  \caption{The van der Waals PMF, $U(r)$, as a function of separation $r$ for a pair of benzene molecules. The results are from all-atom MD simulations with a single pair ($\bigcirc$) and a system with one molecule at the center and six surrounding molecules ($\square$, see Fig.~\ref{fig:benzene_oxygen}(b) for the setup). The lines are guides to the eye.}
  \label{fig:PMF_1_6}
\end{figure}

As an example, Fig.~\ref{fig:PMF_1_6} shows the PMF for a pair of benzene molecules. Two sets of results are included. One is computed from all-atom MD simulations with a single pair of atomic groups, similar to the setup shown in Fig.~\ref{fig:benzene_oxygen}(a). The data show the typical feature of intermolecular interactions, i.e., the force is attractive at large separations and repulsive at short separations, with a well-defined minimum at an equilibrium separation around 5 to 6 \AA.

Using the setup in Fig.~\ref{fig:benzene_oxygen}(a), we have sampled the PMFs for all 15 pairs of atomic groups for the branched polyetherimide at 300 K. For atomic groups carrying different charges but mapped to the same type of coarse-grained beads (with the corresponding different charges), the results show that after the subtraction of Coulomb interactions, the van der Waals PMFs are similar. This is the reason that only 15 PMFs are included in the coarse-grained force field. However, a large discrepancy is observed between the coarse-grained and all-atom model. Under standard conditions for temperature and pressure, the mass density of the branched polyetherimide is 1.209 $\text{g}/\text{cm}^3$ but the density from the coarse-grained model, without any modification as discussed below, is much higher at 1.622 $\text{g}/\text{cm}^3$. This difference reflects a ``softness'' issue of the first version of the coarse-grained model constructed here. That is, pressure is underestimated in this coarse-grained model, leading to a polymer density higher than the target value.\cite{XiaoGuo2016} However, coarse-grained models parameterized from all-atom models using a canonical ensemble usually tend to overestimate pressure as the volume-dependence of the many-body PMF is not captured.\cite{WangJunghans2009, Das2010, Dunn2015} As explained below, the ``softness'' issue here lies in our gas-phase sampling scheme for PMF calculations.

We can illustrate the ``softness'' issue more clearly with benzene. Under standard conditions, the mass density of benzene at 300 K is 0.795 $\text{g}/\text{cm}^3$ from all-atom MD simulations, which is only slightly lower than the experimental value of 0.876 $\text{g}/\text{cm}^3$. During coarse-graining, each benzene is grouped into a bead and the van der Waals PMF between two beads serves as the coarse-grained force field. If this PMF is computed with a setup similar to Fig.~\ref{fig:benzene_oxygen}(a) where only a single pair of benzene molecules is utilized, we find that the mass density from the resulting coarse-grained model is much higher at 1.579 $\text{g}/\text{cm}^3$. This soft PMF is shown as circles in Fig.~\ref{fig:PMF_1_6}.

The ``softness'' issue is related to the insufficient sampling of relative configurations at small separations. For a benzene pair at large separations, their conformations are not strongly correlated and an all-atom MD simulation can sufficiently sample all possible configurations. However, when they get close, the two benzene molecules prefer to be in the T-shaped or parallel-displaced configurations, which are energetically favored.\cite{Sinnokrot2002} In the PMF calculation with one pair of benzene molecules at small separations, the contributions of these configurations dominate. However, in a real benzene system, the local packing of two benzene molecules is affected by other surrounding molecules and cannot all assume the lowest-energy configurations. As a result, the average separation between adjacent benzene molecules is larger than the separation at which the PMF calculated with a single pair reaches its minimum. Furthermore, in a polymeric material containing aromatic rings, the rings are connected to other atomic groups. The T-shaped or parallel-displaced configurations are still favored by the aromatic rings but are subjected to the constraints set by the presence of other groups. As a result, a pair of aromatic rings cannot be as close as in the situation where only the two rings are present. In this sense, the ``softness'' issue is the outcome of using single pairs to sample the many-body interactions among atomic groups in a bulk material.

To overcome the ``softness'' problem, we resort to the setup illustrated in Fig.~\ref{fig:benzene_oxygen}(b). To compute the PMF between a pair of atomic groups, we place one group at the center with its center of mass fixed at the origin of a Cartesian coordinate system ($xyz$), replicate the other group six times, and place the six groups around the central group on the $x$, $-x$, $y$, $-y$, $z$, and $-z$ axes with their centers of mass at an equal distance from the origin. In each snapshot, there are therefore six possible configurations between the central group and the surrounding groups, i.e., there are six pairs simultaneously but all are in different configurations. In this approach, the sampling of unfavorable and rare configurations of the pair is enhanced and the system is more suitable to capture the many-body nature of the nonbonded interactions between the atomic groups in the full-atom model being coarse-grained. For the benzene-benzene pair, the PMF calculated with this six-pair geometry is included in Fig.~\ref{fig:PMF_1_6}, which is obviously more repulsive at short distances compared with the PMF calculated with only one pair. The location where the potential reaches its minimum from the six-pair setup has also shifted to a larger value. As a result, the density of benzene from the coarse-grained model based on the nonbonded PMFs calculated with the six-pair geometry is reduced to 0.914 $\text{g}/\text{cm}^3$, much closer to the result of the atomistic model.

If more simultaneous pairs are used in the PMF calculations, we expect the resulting coarse-grained nonbonded potentials to become even more repulsive at short separations and the corresponding polymer density from the coarse-grained model to be reduced further. We have tested a setup in which twelve groups of atoms, all being replicates of the same group, are placed around the central group. The placement is similar to the arrangement of twelve nearest neighbors around the central atom in a face-centered-cubic crystal. Indeed, the resulting PMF is more repulsive at small distances. For benzene, the density from the corresponding PMF is reduced to 0.692 $\text{g}/\text{cm}^3$, which is smaller than the density from the atomistic model. Later on, we will show that the twelve-pair setup makes it harder to introduce a uniform correction term to the coarse-grained force field for the branched polyetherimide. Therefore, we settle with the six-pair setup for PMF calculations. It should also be pointed out that for atomic groups with shapes resembling spheres, the results from the one-pair and six-pair setup are very close. In this regard, the benzene molecule, which has a planar ring shape, is an ideal model system illustrating the difficulty of developing coarse-grained models for molecular and polymeric systems.

\begin{figure}[htb]
  \centering
  \includegraphics[width=0.4\textwidth]{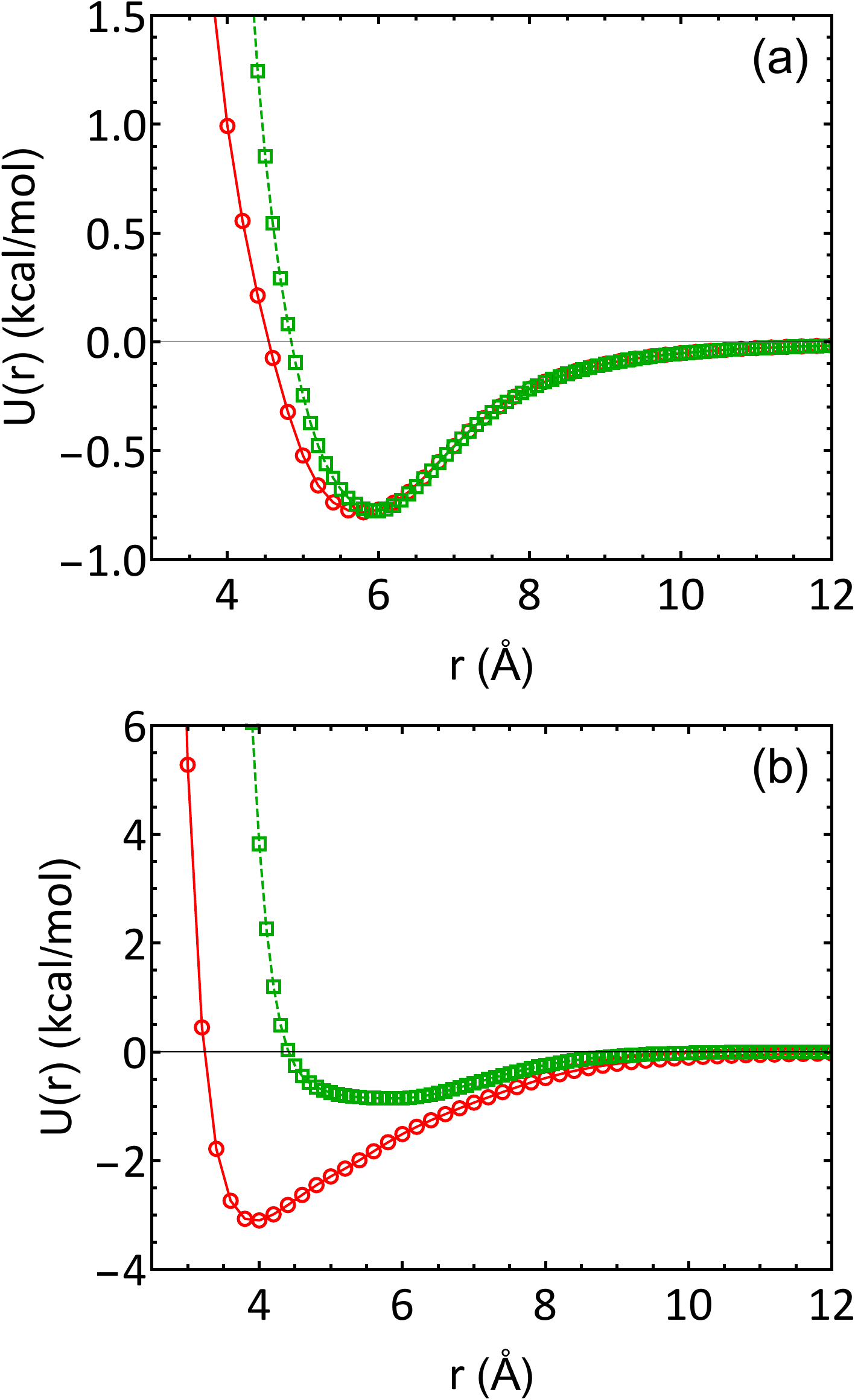}
  \caption{The van der Waals PMF, $U(r)$, as a function of separation $r$ for (a) the D-D pair and (b) the A-D pair. The types of coarse-grained beads are defined in Fig.~\ref{fig:CG_map}. The results are from all-atom MD simulations with a single pair ($\bigcirc$) and a six-pair setup ($\square$). The lines are guides to the eye.}
  \label{fig:PMF_DD_AD}
\end{figure}

For the 5 types of coarse-grained beads defined in Fig.~\ref{fig:CG_map} for the branched polyetherimide, we have performed PMF calculations for all the 15 pairs. Two examples, for the D-D and A-D pair, are shown in Fig.~\ref{fig:PMF_DD_AD}. As expected, the results for the D-D pair are quite similar to those for the benzene pair since the D bead represents an aromatic ring. The PMF from the six-pair setup is slightly more repulsive than that from the one-pair setup at short separations. However, for the A-D pair, the correction introduced by the six-pair setup is quite significant. The location of the PMF minimum shifts from about 4 \AA ~to about 6 \AA, which indicates that the PMF from the six-pair setup is much more repulsive than that from the one-pair setup when the two corresponding atomic groups approach each other. The PMFs for other pairs also show shifts comparably as those in Figs.~\ref{fig:PMF_DD_AD}(a) and (b).

\subsubsection{Corrections to Coarse-Grained Nonbonded Interactions}

The six-pair setup discussed above has improved the sampling of PMFs and enabled the resulting coarse-grained force field to better capture the many-body nature of the nonbonded interactions between atomic groups. However, the PMFs are still too soft at small separations, leading to a polymer density higher than that from the atomistic model. The underlying reason is that even with the six-pair setup, energetically unfavorable configurations are still undersampled at small separations.

In a sampling scheme where only the center-of-mass distance between the two atomic groups being parameterized is constrained but the vector connecting the two centers of mass can still rotate in space, the sampled force on each group contains a contribution from the kinetic entropy associated with two masses rotating at a fixed distance. In three dimensions, this term is $2k_\text{B}T/s$, where $s$ is the center-of-mass separation.\cite{hess2006modeling, Fritz2009, villa2010transferability, Rzepiela2011} In our sampling scheme shown in Fig.~\ref{fig:benzene_oxygen}(b), the vector connecting the center of mass of the central group and that of any surrounding group is completely fixed when the force is sampled at a given value of $s$, because each center of mass is separately fixed. A PMF sampled using the setup in Fig.~\ref{fig:benzene_oxygen}(b) therefore does not contain the kinetic entropic term. However, to compensate for the error in PMFs caused by the undersampling of unfavorable configurations at short ranges, we add a correction term to the nonbonded, van der Waals PMF sampled using the six-pair setup. This term has a form similar to the kinetic entropic term\cite{hess2006modeling, Fritz2009, villa2010transferability, Rzepiela2011} but is slightly modified as
\begin{align} \label{eq:ent_corr_alpha}
\delta \langle f_c \rangle_r = k_\text{B}T\frac{\alpha}{r}~,
\end{align}
where $\alpha$ is treated as a fitting parameter that can be tuned to render the coarse-grained force field to better match the atomistic one. Although the correction term in Eq.~(\ref{eq:ent_corr_alpha}) is purely ad hoc, several considerations lead to the adoption of its particular form. The kinetic entropy of two mass rotating at a fixed separation leads to an effective repulsion between the masses, which to some extent resembles the undersampled repulsion between two atomic groups. Furthermore, the correction should be significant only at small distances where undersampling becomes an issue. The functional form in Eq.~(\ref{eq:ent_corr_alpha}) satisfies this requirement. Finally, even though the correction term expressed in Eq.~(\ref{eq:ent_corr_alpha}) contains temperature, we always parameterize a coarse-grained force field at one temperature and then use it for all temperatures in order to test its temperature transferability. In this sense, $\alpha k_\text{B}T$ in Eq.~(\ref{eq:ent_corr_alpha}) is just a tunable energy factor expressed in the unit of $k_\text{B}T$.

With the force correction in Eq.~(\ref{eq:ent_corr_alpha}), the nonbonded, van der Waals PMF becomes
\begin{align} \label{eq:ent_corr_alpha_PMF}
U(r) & = -\int_{r_m}^r \left[ \langle f_c \rangle_s +\frac{\alpha k_\text{B}T}{s}\right] ds \nonumber \\
& = -\int_{r_m}^r \langle f_c \rangle_s ds - \alpha k_\text{B}T \ln \left( r/r_m \right)~.
\end{align}
Hereafter, a coarse-grained force field based on Eqs.~(\ref{eq:ent_corr_alpha}) and (\ref{eq:ent_corr_alpha_PMF}) is termed a $\text{CG}_\alpha$ model.

To find the optimized value of $\alpha$, we built an atomistic system consisting of 64 branched polyetherimide chains using MAPs.\cite{maps} The system was first heated to 800K and relaxed at that temperature for 5 ns. Then the system was cooled down to 300K within 5 ns. During these steps, the pressure of the system was kept to be one atmosphere. The configuration of the system during the cooling process at a given temperature (between 300 K and 800 K) was taken as a starting state for computing the density of the branched polyetherimide. At each temperature, the system was first relaxed for $3$ ns and its density was calculated in the subsequent $2$ ns using an NPT ensemble. For the coarse-grained system, the same protocol was followed but the system consisted of 1000 branched polyetherimide chains and the time step of the MD simulations was set to 2 fs. This time step is relatively small for coarse-grained MD simulations because the ether oxygen atom has to be kept as a separate bead in the coarse-grained model described here, as explained at the end of Sec.~\ref{sec:CG_mapping}.

\begin{figure}[htb]
  \begin{center}
  \includegraphics[width=0.4\textwidth]{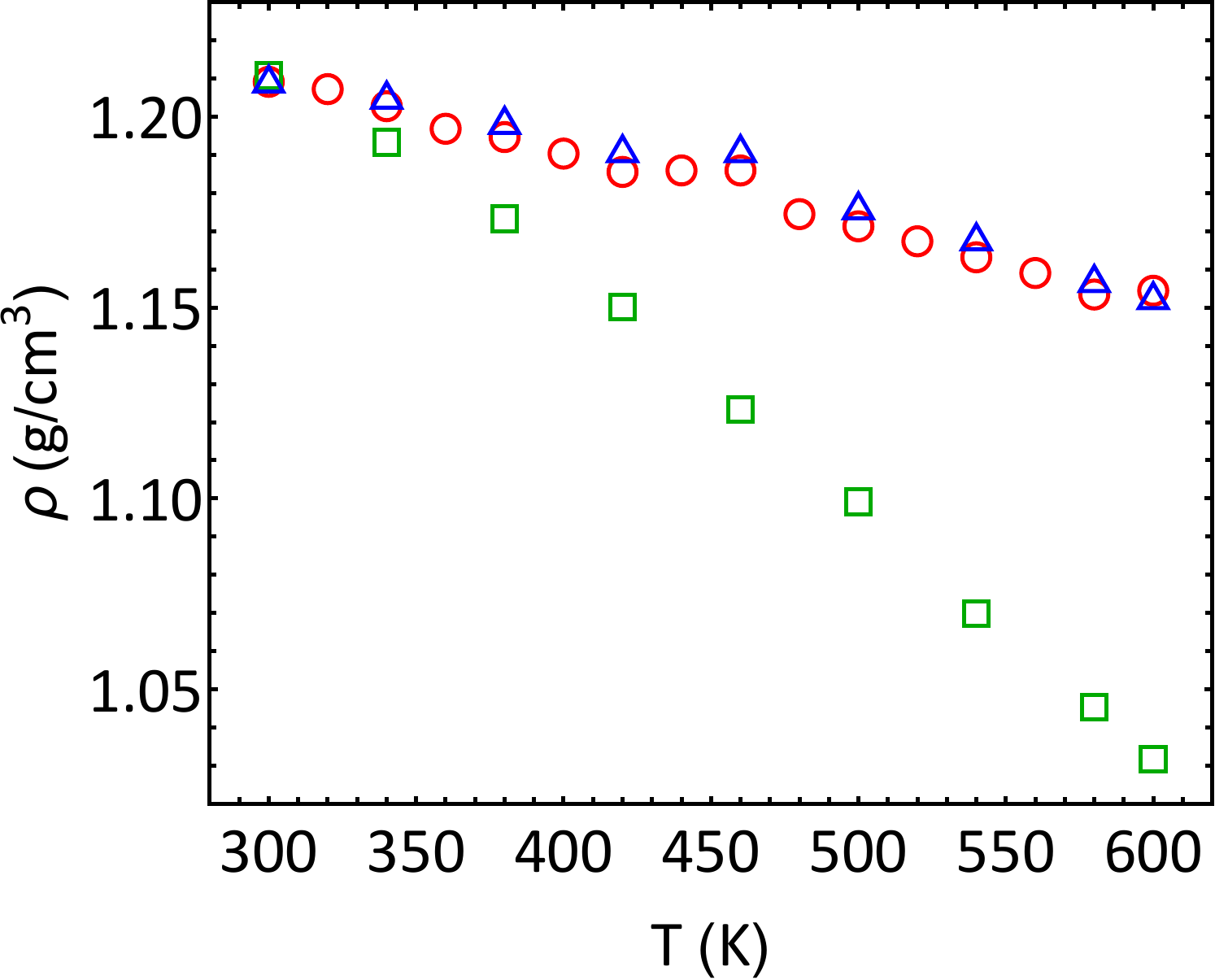}
  \end{center}
  \caption{Density ($\rho$) of the branched polyetherimide as a function of temperature ($T$) from all-atom MD simulations ($\bigcirc$), the $\text{CG}_\alpha$ model ($\square$), and the $\text{CG}_\alpha^n$ model ($\triangle$).}
  \label{fig:den_PEI}
\end{figure}

We find that a single value of $\alpha$ is sufficient to cause the density of the branched polyetherimide from the $\text{CG}_\alpha$ model to match that from all-atom MD simulations at a given $T$. For example, $\alpha = 1.97$ for $T=300$ K, as shown in Fig.~\ref{fig:den_PEI}. The $\text{CG}_\alpha$ model based on this value of $\alpha$ is then used for all temperatures. However, when temperature is raised, the $\text{CG}_\alpha$ model predicts a more compressible polymer system than the atomistic model. For example, at $T=600$ K, the $\text{CG}_\alpha$ model gives a density of 1.031 $\text{g}/\text{cm}^3$ for the branched polyetherimide, while the density from the atomistic model at this temperature is higher at 1.150 $\text{g}/\text{cm}^3$. That is, the $\text{CG}_\alpha$ model parameterized at 300 K overestimates the pressure at higher temperatures.

Later we will discuss the results on the mechanical properties of the branched polyetherimide predicted by the $\text{CG}_\alpha$ model, which are significantly different from those from the all-atom model as well as experimental values. In order to improve the transferability of the coarse-grained model, it is effective to modify the correction term to PMFs as
\begin{align} \label{eq:ent_corr_alpha_n}
\delta \langle f_c \rangle_r = k_\text{B}T\frac{\alpha}{r} \left(\frac{r_0}{r}\right)^n~,
\end{align}
where $n$ is an additional fitting parameter and $r_0$ is the separation between a pair of atomic groups at which their mutual force from the PMF calculation reaches minimum (i.e., where the force is most attractive). The rationale behind this modification is that the correction should only come into play at small separations, while the form in Eq.~(\ref{eq:ent_corr_alpha}) grows in magnitude continuously from 0 to $\infty$ as $r$ is reduced from $r_m$ to 0. On the other hand, the form in Eq.~(\ref{eq:ent_corr_alpha_n}) guarantees that as long as $n$ is sufficiently large, the correction is negligible when $r \gg r_0$ and its importance grows in a power law for $r < r_0$. The coarse-grained force field with Eq.~(\ref{eq:ent_corr_alpha_n}) as the correction term is designated as the $\text{CG}_\alpha^n$ model.

Since the $\text{CG}_\alpha^n$ model has two fitting parameters, $\alpha$ and $n$, we can tune the model to have a density match with the atomistic model at two temperatures. As shown in Fig.~\ref{fig:den_PEI}, with $\alpha = 0.395$ and $n = 15$, the density of the branched polyetherimide from the $\text{CG}_\alpha^n$ model matches that from the all-atom model at both $T=300$ K and 600 K. Furthermore, without further tuning the densities at other intermediate temperatures also match between the $\text{CG}_\alpha^n$ and all-atom model. That is, the $\text{CG}_\alpha^n$ model, which is parameterized at one temperature (300 K here) and used for all temperatures, captures the thermal expansion property of the branched polyetherimide encoded in the atomistic model. As pointed out previously, reproducing the thermal expansion coefficient for a fluid system in a coarse-grained model is crucial for its temperature transferability.\cite{Qian2008, Carbone2008, brini2011conditional, Karimi-Varzaneh2012} As discussed in the next section, this will also make the coarse-grained model to better capture the mechanical properties of the target polymer.

In this paper, we employ an ad hoc correction term in Eq.~(\ref{eq:ent_corr_alpha_n}) for the PMFs sampled with gas-phase MD simulations to ensure that the thermal expansion behavior of the branched polyetherimide is captured in the $\text{CG}_\alpha^n$ model. This correction is introduced to address the ``softness'' issue of the PMFs at short ranges that makes the pressure to be underestimated.\cite{Karimi-Varzaneh2012} That is, the correction term is included essentially for ``pressure correction'', which is reminiscent of the linear correction method used by IBI to get the right pressure in a coarse-grained model.\cite{Muller-Plathe2002, Reith2003, WangJunghans2009} In the force-matching coarse-graining approach, pressure correction can also be done in a self-consistent manner.\cite{Lebold2019} Reproducing pressure correctly is crucial for the thermal expansion behavior of a material to be captured as it is defined as a constant pressure. Our results show that the $\text{CG}_\alpha^n$ model, parameterized at one temperature with one set of $(\alpha, n)$ for all the PMFs, is able to get the pressure right in a range of temperatures and thus yields the correct polymer density in that temperature range.

It should be clarified that the six-pair setup for PMF calculations and the correction in either Eq.~(\ref{eq:ent_corr_alpha}) or Eq.~(\ref{eq:ent_corr_alpha_n}) are used as complementary measures to improve the coarse-grained model discussed here. If the six-pair setup was not used, we would need to have a separate $\alpha$ or $(\alpha,n)$ combination for each pair of atomic groups (i.e., for each PMF) in order to achieve a density match (at one temperature for the $\text{CG}_\alpha$ model or a range of temperatures for the $\text{CG}_\alpha^n$ model). The number of fitting parameters would then increase significantly, though they could still be determined through efficient optimization processes such as variational approaches used in the force-matching\cite{Noid2008a} or relative entropy\cite{Shell2008, ChaimovichShell2011} coarse-graining methods. However, with the six-pair setup, just one $\alpha$ or one $(\alpha,n)$ combination is needed for all the PMFs in the $\text{CG}_\alpha$ model or the $\text{CG}_\alpha^n$ model, respectively. The data for all the 15 PMFs based on the $\text{CG}_\alpha^n$ model are included in the Supporting Information.

\section{Applications of the Coarse-Grained Model of the Branched Polyetherimide} \label{sec:CG_model_application}

We have compared the chain dynamics and size distribution in a melt state from the all-atom model and the $\text{CG}_\alpha^n$ model. The results are included in the Supporting Information. The distribution of chain sizes in the two models shows a reasonable agreement. However, at 700 K the chain diffusion is about 11 times faster in the coarse-grained model. This is expected since the coarse-grained potentials are smoother than the interatomic ones. In this section we apply the coarse-grained models developed previously to study the mechanical, structural, and rheological properties of the branched polyetherimide. We show that the $\text{CG}_\alpha^n$ model reasonably captures these properties.

\subsection{Mechanical Moduli}

\begin{figure}[htb]
\centering
\includegraphics[width=0.4\textwidth]{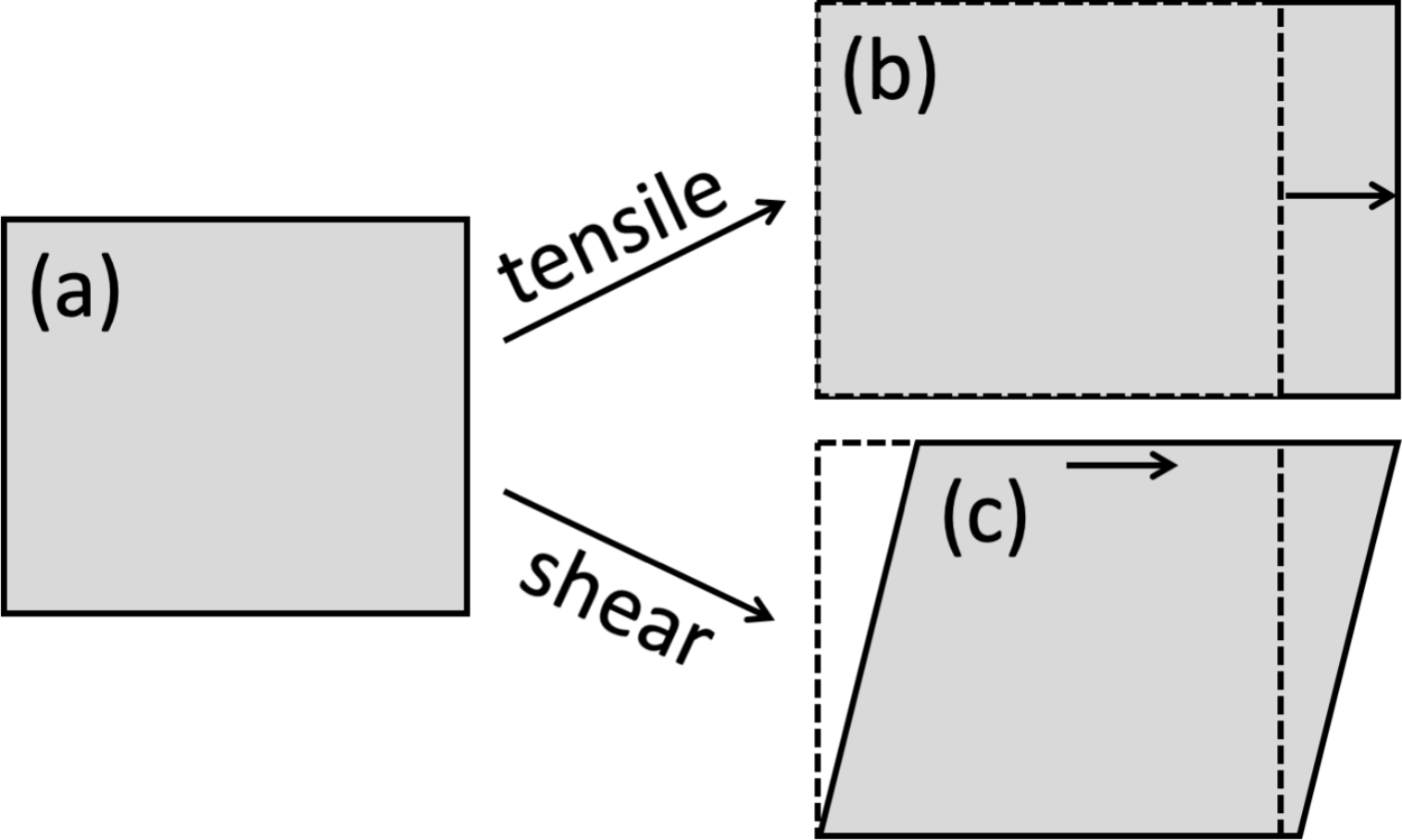}
\caption{Setup of simulations used to compute the mechanical moduli of the branched polyetherimide. The polymer domain undergoes either a tensile deformation [(a)$\rightarrow$(b)] and a simple shear [(a)$\rightarrow$(c)].}
\label{fig:deform}
\end{figure}

We compute the mechanical moduli of the branched polyetherimide with both atomistic and coarse-grained models. The setup of such simulations is shown in Fig.~\ref{fig:deform}. A system of polyetherimide chains (64 atomistic chains or 1000 coarse-grained chains; each chain contains 3 branches with one branch shown in Fig.~\ref{fig:CG_map}) is first equilibrated in an NPT ensemble at 300 K and 1 atmosphere. Periodic boundary conditions are used in all directions. After equilibration, the simulation box size is 62.2 \AA~$\times$ 62.2 \AA~$\times$ 62.2 \AA ~for the atomistic system and 155.7 \AA~$\times$ 155.7 \AA~$\times$ 155.7 \AA ~for the coarse-grained system. An NVT ensemble is used from this point forward and the stress tensor of the equilibrated system is computed as a reference. Then either a tensile or shear strain is applied to deform the simulation box, as shown in Fig.~\ref{fig:deform}. After deformation, the system is relaxed to remove transient effects and the stress tensor under the given strain is computed. The change of the stress tensor is analyzed as a function of the applied strain, which yields the mechanical moduli as well as Poisson's ratio of the materials.

For an isotropic material, its mechanical moduli are determined by only two independent parameters, $\lambda$ and $\mu$, where $\lambda$ is Lam\'{e}'s first parameter and $\mu$ is Lam\'{e}'s second parameter or the shear modulus of the material. The mechanical moduli are then given by
\begin{align}
&K =\lambda+\frac{2}{3}\mu~, \nonumber\\
&G = \mu~, \nonumber\\
&E=\mu\left(\frac{3\lambda+2\mu}{\lambda+\mu}\right)~, \nonumber\\
&\nu = \frac{\lambda}{2(\lambda+\mu)}~,
\end{align}
where $K$ is the bulk modulus, $G$ is the shear modulus, $E$ is Young's modulus, and $\nu$ is Poisson's ratio.

Lam\'{e}'s parameters can be determined by computing the stiffness tensor,
\begin{align}\label{tb:stiffness_tensor}
C=\begin{bmatrix}
\lambda+2\mu & \lambda           & \lambda           & 0 & 0 & 0 \\ 
\lambda           & \lambda+2\mu & \lambda           & 0 & 0 & 0\\ 
\lambda           & \lambda           & \lambda+2\mu & 0 & 0 & 0 \\ 
0 & 0 & 0 & \mu & 0 & 0 \\ 
0 & 0 & 0 & 0 & \mu & 0 \\ 
0 & 0 & 0 & 0 & 0 & \mu
\end{bmatrix}.
\end{align}
The matrix element $C_{ij}$ can be computed from $C_{ij} = \sigma_{i}/\epsilon_{j}$, where $\epsilon_{j}$ is the $j$-th component of the strain tensor and $\sigma_{i}$ is the $i$-th component of the corresponding stress tensor with both tensors expressed in a vector form. The subindices, $i$ and $j$, each running from 1 to 6, denote the $xx$, $yy$, $zz$, $yz$, $xz$, and $xy$ components, respectively, of the strain or stress tensor expressed in a $3\times 3$ matrix form in a three-dimensional Cartesian system. In particular, $\sigma_1 = \sigma_{xx}$, $\sigma_2 = \sigma_{yy}$, $\sigma_3 = \sigma_{zz}$, $\sigma_4 = \sigma_{yz}$, $\sigma_5 = \sigma_{xz}$, and $\sigma_6 = \sigma_{xy}$ for the stress tensor. For the strain tensor, $\epsilon_1 = \epsilon_{xx}$, $\epsilon_2 = \epsilon_{yy}$, $\epsilon_3 = \epsilon_{zz}$, $\epsilon_4 = 2\epsilon_{yz}$, $\epsilon_5 = 2\epsilon_{xz}$, and $\epsilon_6 = 2\epsilon_{xy}$. Therefore, $\epsilon_1$ is a tensile strain along the $x$-axis while $\epsilon_4$ is twice the shear strain applied along the $y$-axis on a surface perpendicular to the $z$-axis, and so on.

We first use tensile deformations to compute the top-left block of the stiffness tensor. Then Lam\'{e}'s parameters are computed via
\begin{align}\label{modulus}
&\lambda = \frac{1}{6}(C_{12}+C_{13}+C_{21}+C_{23}+C_{31}+C_{32})~,\\
&\mu = \frac{1}{6}(C_{11}+C_{22}+C_{33}-3\lambda).
\end{align}
Shear deformations are also simulated to independently determine $\mu$ and the results are consistent with those from tensile deformations.

\begin{table}[htb]
\centering
\caption{Mechanical moduli and Poisson's ratio of the branched polyetherimide at 300 K. The unit of $K$, $G$, $E$, and $\lambda$ is GPa. The row of $\nu$ and the column for the parameter $w$ are dimensionless. The experimental data were from the measurements conducted by Roy Odle at SABIC.}
\label{tb:moduli_300K}
\begin{tabular}{|c|c|c|c|c|c|}
\hline
                  & Atomistic & $\text{CG}_\alpha$ & $\text{CG}_\alpha^n$ & Experimental & $w$   \\
\hline
$K$          & 4.185 & 0.727 & 2.110 & 4.297--4.942 &0.50 \\
\hline
$G$          & 0.770 & 0.120 & 0.469 & 1.059-1.070 & 0.61 \\
\hline
$E$           & 2.178 & 0.342  & 1.311 & 2.965 & 0.60 \\
\hline
$\lambda$ & 3.671 & 0.647  & 1.797 & 3.584-4.236 & 0.49 \\
\hline
$\nu$         & 0.413 & 0.422  & 0.397 & 0.385-0.400 & \\
\hline
\end{tabular}
\end{table}

\begin{table}[htb]
\centering
\caption{Mechanical moduli and Poisson's ratio of the branched polyetherimide at 400 K. The unit of $K$, $G$, $E$, and $\lambda$ is GPa. The row of $\nu$ and the column for the parameter $w$ are dimensionless.}
\label{tb:moduli_400K}
\begin{tabular}{|c|c|c|c|c|}
\hline
                & Atomistic & $\text{CG}_\alpha$ & $\text{CG}_\alpha^n$ & $w$\\
\hline
$K$          & 3.754 & 0.345   & 1.724 & 0.46 \\
\hline
$G$          & 0.653 & 0.0167 & 0.310 & 0.47 \\
\hline
$E$           & 1.852 & 0.0491 &0.879 & 0.47 \\
\hline
$\lambda$ & 3.319 & 0.334  & 1.517 & 0.46 \\
\hline
$\nu$         & 0.418 & 0.476   & 0.415 & \\
\hline
\end{tabular}
\end{table}

The results on the mechanical moduli of the branched polyetherimide are computed with the atomistic, $\text{CG}_\alpha$, and $\text{CG}_\alpha^n$ models and are summarized in Table~\ref{tb:moduli_300K} for $T=300$ K and in Table~\ref{tb:moduli_400K} for $T=400$ K. The experimental values at $T=300$ K are also included in Table~\ref{tb:moduli_300K}. All models yield very good results on Poisson's ratio that match with the experimental value. With regard to the mechanical moduli, the data further show that the results from the atomistic model are close to the experimental ones and the $\text{CG}_\alpha^n$ model is significantly improved compared with the $\text{CG}_\alpha$ model in terms of matching the atomistic model. It is also interesting to notice that if we define $w$ as the ratio between the value of a mechanical modulus from the $\text{CG}_\alpha^n$ model and that from the atomistic model, then $w$ is around 0.5 for all mechanical moduli at either $T=300$ K or 400 K. This comparison indicates that the $\text{CG}_\alpha^n$ model developed in this paper is able to capture the mechanical properties of the branched polyetherimide with an almost constant scaling factor about 0.5. In this sense, the $\text{CG}_\alpha^n$ model is transferable temperature-wise. Nevertheless, such scaling is often used in coarse-grained models of molecular and polymeric systems as dynamics is speed up when the number of degrees of freedom is reduced and potentials are smoothed out during coarse-graining.\cite{Karimi-Varzaneh2012}

The much improved performance, including temperature transferability, of the $\text{CG}_\alpha^n$ model is due to the fact that it captures the thermal expansion property of the branched polyetherimide predicted by the atomistic model.\cite{Qian2008, Carbone2008, brini2011conditional, Karimi-Varzaneh2012} This behavior can be physically explained with the Gr\"{u}neisen law that uses a parameter, $\gamma$, to describe the effect of a changing temperature on the size and dynamics of a crystal lattice. One expression of $\gamma$ is
\begin{equation}\label{eq:gruneisen}
\gamma=\frac{\alpha_V K }{C_V \rho}~,
\end{equation}
where $\alpha_V$ is the volumetric thermal expansion coefficient and $C_V$ is the constant-volume specific heat of the crystal. The physical implication of the Gr\"{u}neisen law is that the thermal expansion behavior of a crystal, or more generally a solid, is intrinsically connected to its mechanical properties.\cite{Barker1963, Ledbetter1991, LeBourhis2007, Sanditov2014, Laplanche2018} If we assume the Gr\"{u}neisen law also applies to the branched polyetherimide, then the law indicates that the ratio $\gamma C_V/K$ should be the same for the atomistic and $\text{CG}_\alpha^n$ model as they yield matching polymer densities as well as volumetric thermal expansion coefficients. The fact that the value of $K$ from the $\text{CG}_\alpha^n$ model is about 50\% of that from the atomistic model thus indicates that $\gamma C_V$ should scale similarly between the two models.

\subsection{Pair Correlation Functions}

\begin{figure*}[htb]
 \centering
  \includegraphics[width=\textwidth]{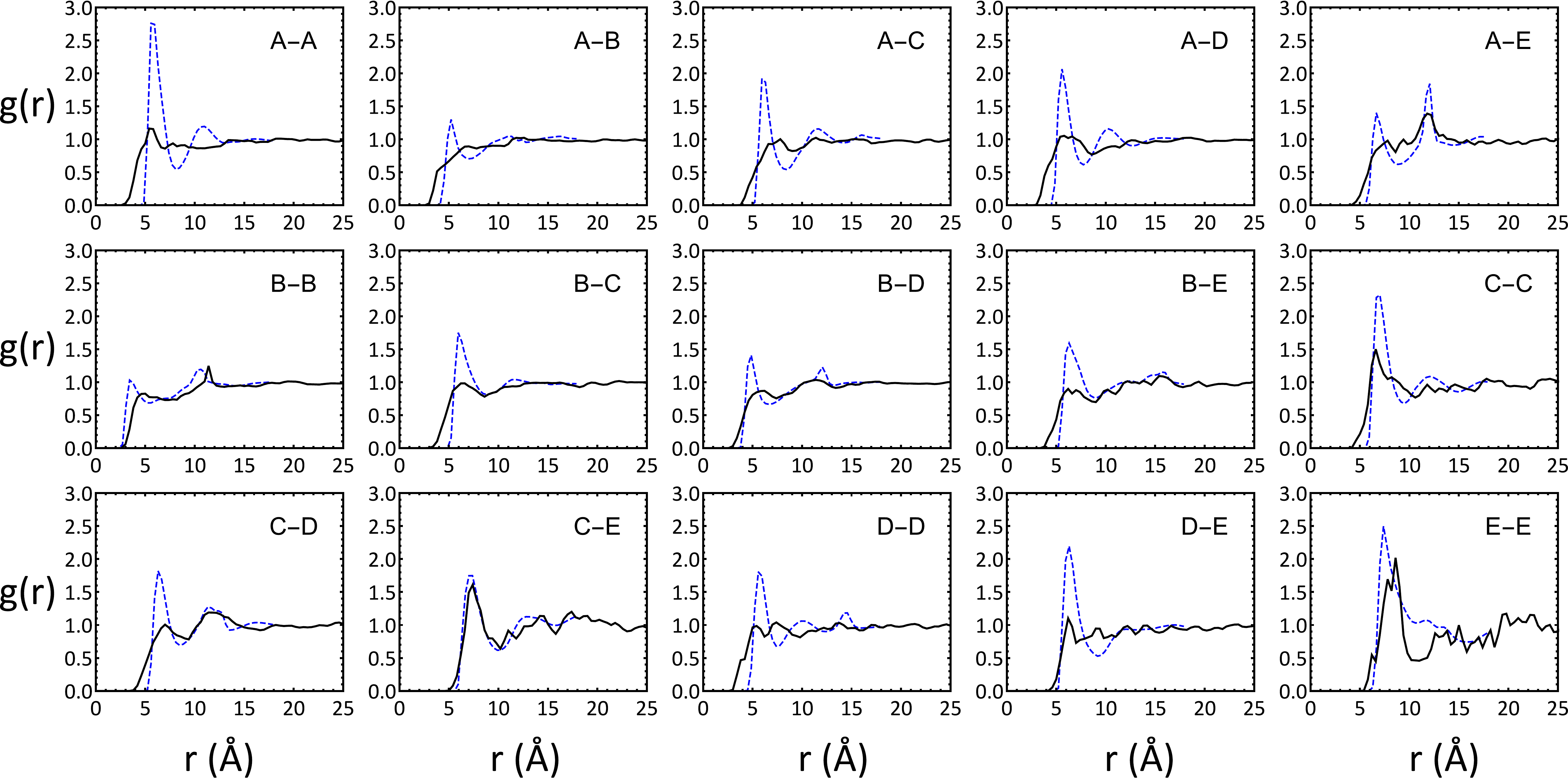}
  \caption{Comparison of pair correlation functions, $g(r)$, from the atomistic model (black solid lines) and the $\text{CG}_\alpha^n$ model (blue dashed lines) for all pairs of atomic groups or corresponding coarse-grained beads. The types of coarse-grained beads are defined in Fig.~\ref{fig:CG_map}.}
  \label{fig:GofR}
\end{figure*}

We have computed the pair correlation functions, $g(r)$, of all the 15 pairs of the atomic groups defined in Fig.~\ref{fig:CG_map} with all-atom MD simulations and those of the corresponding coarse-grained beads with the $\text{CG}_\alpha^n$ model. The comparison is shown in Fig.~\ref{fig:GofR}. The locations of peaks in $g(r)$ generally match reasonably but their heights differ significantly. The results of $g(r)$ from the atomistic model indicate that the arrangement of the atomic groups in the branched polyetherimide is rather structureless, particularly beyond the first peak of $g(r)$. On the other hand, the coarse-grained beads show more local ordering as evidenced by a strong first peak in $g(r)$ for almost all the pairs. Beyond the first peak, the pair correlation functions of the coarse-grained beads match reasonably well with those of the corresponding atomic groups. The results in Fig.~\ref{fig:GofR} are not surprising as $g(r)$ never enters the process when the coarse-grained model is constructed. The discrepancy, which is acceptable from our perspective, is the price that has to be paid since our goal is to make the coarse-grained model transferable and easily expandable. We obtained the PMFs among the coarse-grained beads through separate gas-phase MD calculations for each pair of atomic groups. Although the sampling of PMF is speed up by using the six-pair setup and a correction term is introduced to enable the coarse-grained model based on the resulting PMFs to be temperature transferable, these PMFs are still poor approximations of the true many-body PMFs that can be systematically derived by rigorously following the force-matching coarse-graining procedure of Voth, Andersen, Noid, and coworkers.\cite{Noid2008a} That is, the PMF of a pair of atomic groups needs to be computed in the presence of all other atomic groups for the structure of a molecular system to be accurately preserved in the corresponding coarse-grained model.

\subsection{Shear Rheology}

\begin{figure}[htb]
 \centering
  \includegraphics[width=0.4\textwidth]{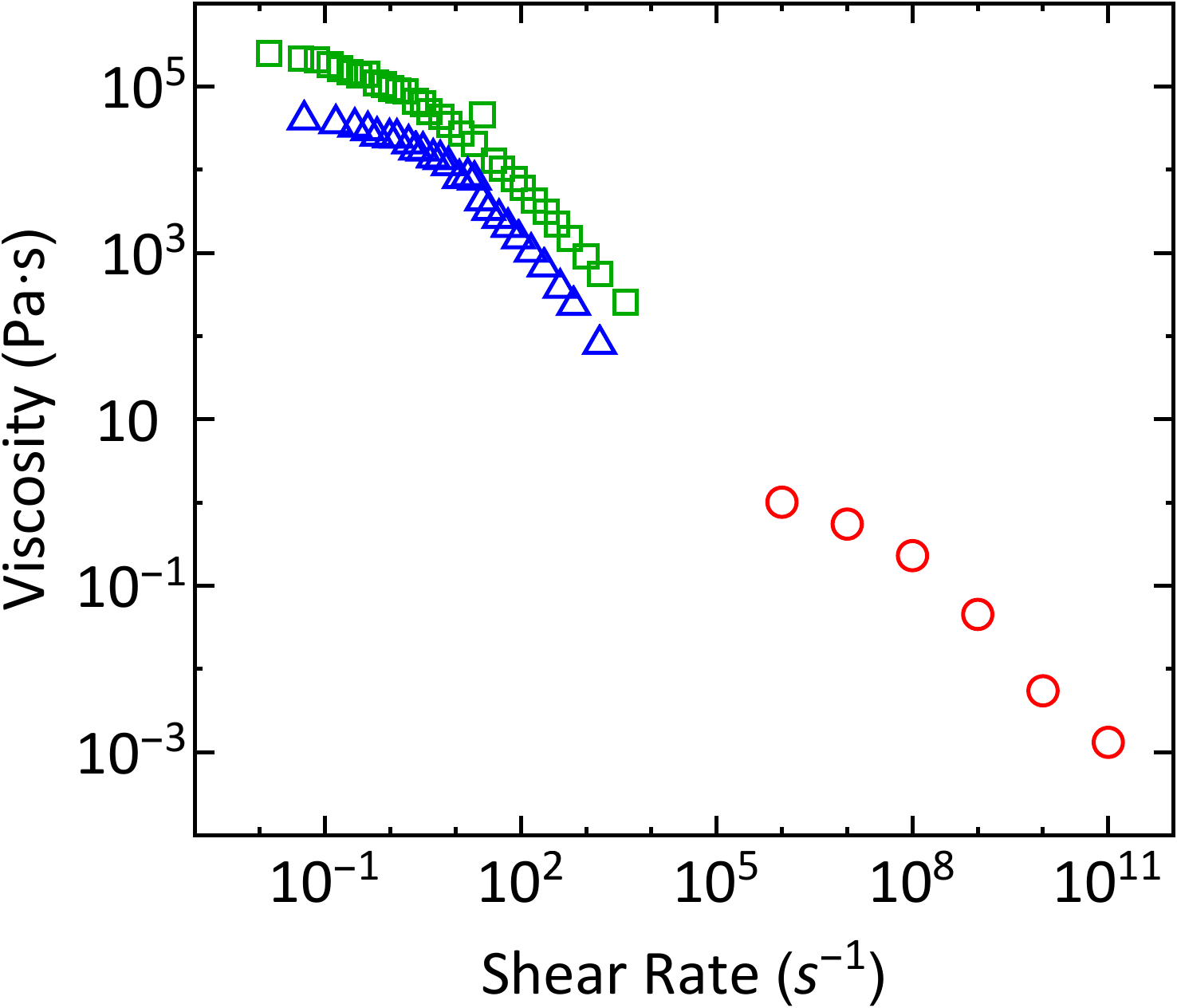}
  \caption{Shear viscosity as a function of shear rate from experiments ($\triangle$ for a branched polyetherimide with $M_n = 18.8$ kDa and $\square$ for a linear polyetherimide with $M_n = 20$ kDa; experimental data by courtesy of Timothy E. Long and Josh Wolfgang at Virginia Tech) and MD simulations ($\bigcirc$ for a branched polyetherimide with $M_n = 2.74$ kDa). For all systems, $T=563$ K.}
  \label{fig:shear_viscosity}
\end{figure}

Finally, we apply the $\text{CG}_\alpha^n$ model discussed above to study the rheological properties of the branched polyetherimide. The simple-shear setup shown in Fig.~\ref{fig:deform} was used to compute the viscosity of the polymer in a triclinic simulation box under a given shear rate. The system consisted of 8000 branched polyetherimide chains with $M_n = 2.74$ kDa dispersed in a cubic simulation box with a side length of 315.38 \AA. The temperature was fixed at $T=563$ K, at which the branched polyetherimide formed a melt. This was also the temperature at which the experimental data on viscosity were obtained. At a fixed shear rate, the shear stress in the melt was computed in MD simulations with the $\text{CG}_\alpha^n$ force field and used to determine the viscosity. The MD data for shear rates ranging from $10^6~\text{s}^{-1}$ to $10^{11}~\text{s}^{-1}$ are shown in Fig.~\ref{fig:shear_viscosity}, together with the experimental data for a branched polyetherimide with $M_n = 18.8$ kDa and a linear polyetherimide with $M_n = 20$ kDa. The shear thinning behavior is obvious from both MD and experimental results. Furthermore, the upper range of the experimentally-probed shear rates is about $10^4~\text{s}^{-1}$ to $10^5~\text{s}^{-1}$. The $\text{CG}_\alpha^n$ model enables us to approach the experimental range of shear rates as well as the range of low shear rates where the Newtonian plateau occurs. However, more work is still needed to close the gap.

\section{Conclusions} \label{sec:conclusion}

We have developed a coarse-grained model of a branched polyetherimide derived from Ultem on the basis of chemistry-informed grouping of atoms, parameterization of bond and angle interactions by direct Boltzmann inversion, and parameterization of nonbonded interactions via potential of mean force (PMF) calculations with gas-phase all-atom molecular dynamics simulations of atomic group pairs. Our results show that a six-pair setup, in which one atomic group is placed at the origin and six replicates of another atomic group are placed around the central group in a NaCl structure, can be used to speed up and improve the sampling of PMFs. An ad hoc correction is added to the PMFs in order for the coarse-grained model to correctly reproduce the thermal expansion property of the polymer. This correction is reminiscent of the potential correction used in the iterative Boltzmann inversion method,\cite{Muller-Plathe2002, Reith2003, WangJunghans2009} as well as the ``volume force'' in the variational force-matching approach.\cite{Lebold2019} All these corrections are included to make the pressure right in the coarse-grained models. After the correction, our coarse-grained model is able to reproduce the polymer density in a range of temperatures and therefore its thermal expansion coefficient, compared to the all-atom model. This turns out to be the key to making the corresponding coarse-grained model transferable temperature-wise. As a result, the coarse-grained model is able to reproduce the mechanical moduli of the branched polyetherimide at different temperatures within a (temperature-independent) constant scaling factor, which is around 0.5 here. The underlying physical reason is that a solid's mechanical properties are strongly correlated to its thermal expansion coefficient.\cite{Barker1963, Ledbetter1991, LeBourhis2007, Sanditov2014, Laplanche2018} The coarse-grained model further enables us to approach the range of shear rates accessible to rheology experiments and probe the polymer's rheological behavior such as shear thinning. The coarse-grained model only fairly captures the structural property of the polymer and future improvements are still needed in this respect.

\begin{acknowledgments}
This article is based on the results from work supported by SABIC. The authors gratefully acknowledge many stimulating discussions with Dr. Timothy E. Long and Dr. Guoliang (Greg) Liu. The authors thank Dr. Timothy E. Long and Mr. Josh Wolfgang for providing the experimental data in Fig.~\ref{fig:shear_viscosity}. The authors acknowledge Advanced Research Computing at Virginia Tech (URL: http://www.arc.vt.edu) for providing computational resources and technical support that have contributed to the results reported within this article. The authors also gratefully acknowledge the support of NVIDIA Corporation with the donation of the Tesla K40 GPUs used for this research.
\end{acknowledgments}




\clearpage
\newpage
\onecolumngrid
\renewcommand{\thefigure}{S\arabic{figure}}
\setcounter{figure}{0}
\renewcommand{\theequation}{S\arabic{equation}}
\setcounter{equation}{0} 
\renewcommand{\thepage}{SI-\arabic{page}}
\setcounter{page}{1}    
\begin{center}
{\bf SUPPORTING INFORMATION}
\end{center}


\noindent \textbf{S1. Equivalence between the ``Recentering'' and Constraint-Force Schemes of Fixing a Center of Mass}

\noindent Here we prove the equivalence between the ``recentering'' and constraint-force schemes of fixing the center of mass of a group of atoms. We first examine the movement of each atom in the ``recentering'' approach. We use $\vec{f}_i$ to denote the total force on the $i$-th atom in a group from its interactions with all other atoms in the system. Then after an infinitesimal time $dt$, the velocity of this atom becomes
\begin{align}
\vec{v}_i (t+dt) = \vec{v}_i + \frac{\vec{f}_i}{m_i} dt~,
\end{align}
and the displacement of this atom is
\begin{align}
d\vec{r}_i = \vec{v}_i dt + \frac{1}{2} \frac{\vec{f}_i}{m_i} (dt)^2~,
\end{align}
where $\vec{v}_i$ is the velocity of the $i$-th atom at time $t$, and $m_i$ is its mass. To fix the center of mass, we require $\vec{v}_i$ to satisfy $\sum_i m_i \vec{v}_i = 0$, where the summation is over all atoms in the group under consideration. The displacement of the group's center of mass before ``recentering'' is therefore
 \begin{align}
d\vec{R} & \equiv  \frac{ \sum_i m_i d\vec{r}_i }{\sum_i m_i} \nonumber \\
& =  \frac{ \sum_i m_i \vec{v}_i + \frac{1}{2} dt \sum_i \vec{f}_i }{\sum_i m_i} dt \nonumber \\
& = \frac{1}{2}  \frac{\sum_i \vec{f}_i}{\sum_i m_i} (dt)^2 .
\end{align}
During ``recentering'', each atom in the group is displaced by $-d\vec{R}$ to move the group's center of mass back to its starting location and the renormalized displacement of the $i$-th atom becomes
\begin{align}
d\vec{r}_{i, R} = d\vec{r}_i - d\vec{R} =  \vec{v}_i dt + \frac{1}{2} \frac{\vec{f}_i}{m_i} (dt)^2 - \frac{1}{2}  \frac{\sum_i \vec{f}_i}{\sum_i m_i} (dt)^2 ~.
\end{align}
It is easy to prove that $\sum_i (m_i \times d\vec{r}_{i, R}) = 0$, indicating that the center of mass is fixed. The velocity of the center of mass, $\vec{v}_c$, before ``recentering'' is
 \begin{align}
\vec{v}_c = \frac{\sum_i \vec{f}_i}{\sum_i m_i} dt~.
\end{align}
When this velocity is subtracted from the velocity of each atom in the group, the renormalized velocity of the $i$-th atom becomes
\begin{align}
\vec{v}_{i, R}(t+dt) =\vec{v}_i(t+dt) - \vec{v}_c  = \vec{v}_i + \frac{\vec{f}_i}{m_i} dt -  \frac{\sum_i \vec{f}_i}{\sum_i m_i} dt.
\end{align}
The velocity of the group's center of mass after ``recentering'' is reduced to zero, i.e., $\sum_i [m_i \times \vec{v}_{i, R}(t+dt)]=0$. As expected, ``recentering'' renders the center of mass of the group fixed.

Next, we consider the constraint-force approach by applying an extra constraining force, $\vec{f}_{i,C}$, to the $i$-th atom in the group. The velocity of the $i$-th atom after an infinitesimal time $dt$ is
\begin{align}
\vec{v}_{i,C} (t+dt) = \vec{v}_i + \frac{\vec{f}_i + \vec{f}_{i,C}}{m_i} dt~,
\end{align}
and its corresponding displacement is
\begin{align}
d\vec{r}_{i, C} = \vec{v}_i dt + \frac{1}{2} \frac{\vec{f}_i +  \vec{f}_{i,C}}{m_i} (dt)^2~.
\end{align}
It is easy to show that if
\begin{align}
\vec{f}_{i,C} \equiv -m_i \frac{\sum_j \vec{f}_j}{\sum_j m_j}~,
\end{align}
then
\begin{align}
d\vec{r}_{i, C} = d\vec{r}_{i, R}~~~\text{and}~~~\vec{v}_{i,C} (t+dt)=\vec{v}_{i,R} (t+dt) ~.
\end{align}
This proves that the ``recentering'' and constraint-force schemes are equivalent. Furthermore, it can be noted that
\begin{align}\label{eq:cf_1}
\sum_i \vec{f}_{i,C} = - \sum_i \vec{f}_i~,
\end{align}
and
\begin{align}\label{eq:cf_2}
\sum_i \vec{r}_i \times \vec{f}_{i,C} = 0~,
\end{align}
where $\vec{r}_i$ is the position vector of the $i$-th atom relative to the group's center of mass. Eqs.~(\ref{eq:cf_1}) and (\ref{eq:cf_2}) are two natural requirements of the constraint-force scheme of fixing a center of mass. That is, the total force from the constraints should balance the total force exerted on all atoms in the group by other atoms in the system, which makes the acceleration of the center of mass to be zero. When the initial velocity of the center of mass is zero, it is naturally fixed. Furthermore, the constraint forces should have zero torque on the group to which they are applied. The rotation of the group around its center of mass is purely determined by the interactions with atoms in other groups.


\begin{figure}[htb]
  \centering
  \includegraphics[width = 0.6\textwidth]{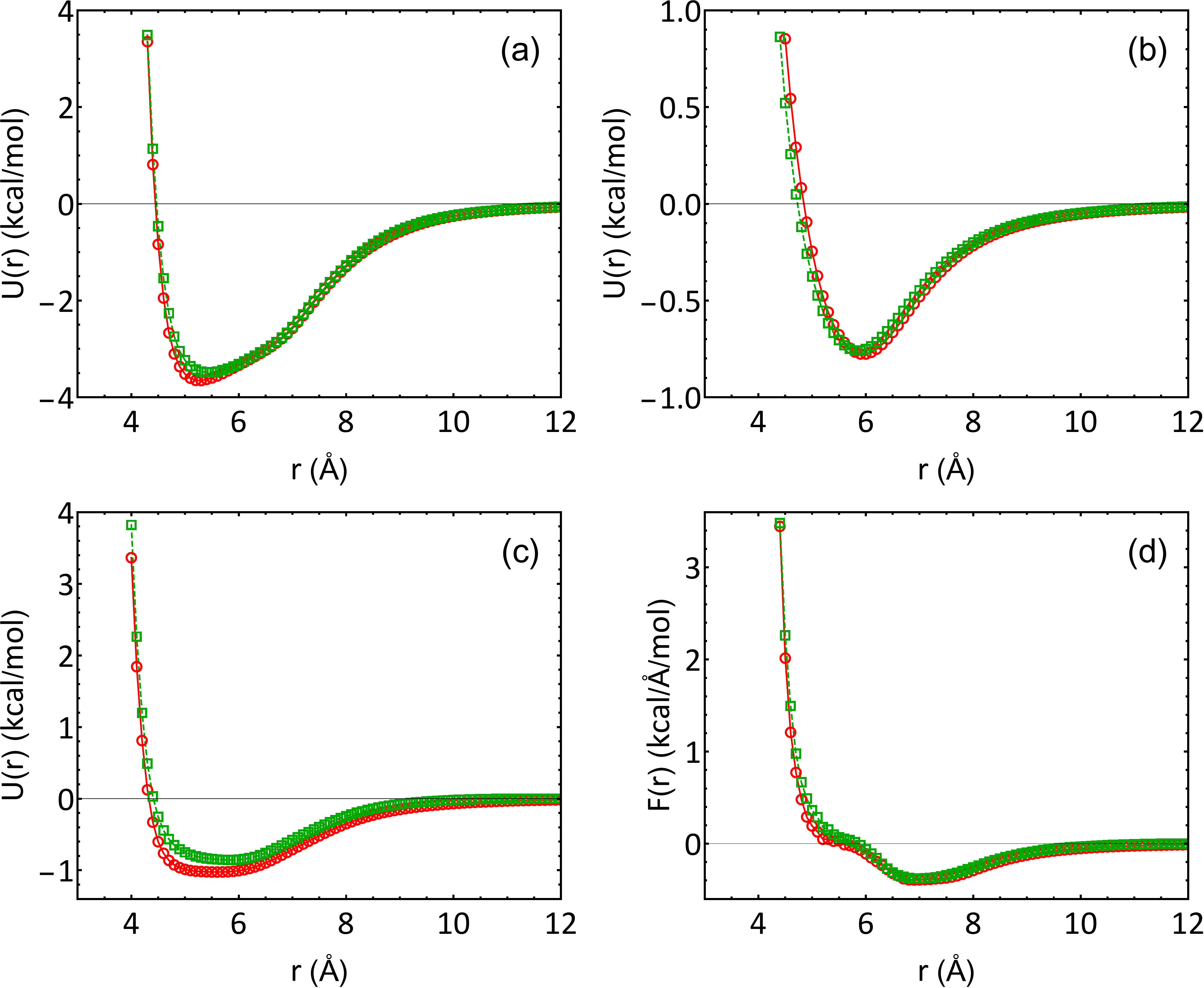}
  \caption{The van der Waals PMF, $U(r)$, as a function of separation $r$, (a) for a pair of \#1 atomic groups (circles) and a pair of \#3 atomic groups (squares); (b) for a pair of \#2 atomic groups (circles) and a pair of \#8 atomic groups (squares); (c) between \#1 atomic group and \#8 atomic group (circles) and between \#3 atomic group and \#2 atomic group (squares). In (d), the mean force, $F(r)$, between \#1 atomic group and \#8 atomic group (circles) and between \#3 atomic group and \#2 atomic group (squares) is plotted against $r$. The atomic groups are defined in Table I of the main text.}
  \label{fg:PMF_comparison}
\end{figure}

\noindent \textbf{S2. Additional Results on the Potential of Mean Force }

\noindent Here we include additional results on the potential of mean force (PMF) among atomic groups mapped to the same coarse-grained beads in terms of van der Waals interactions but carrying different charges. As discussed in the main text, atomic group 1 is mapped to the same coarse-grained bead as groups 3 and 7 in terms of pairwise van der Waals interactions but the former carries a different charge from the latter two and a slightly higher mass (by $\sim 0.7$\%) as well. Similarly, atomic groups 2 and 8, mapped to the same coarse-grained beads, have the same mass but different charges. In Fig.~\ref{fg:PMF_comparison}, the van der Waals PMF between a pair of \#1 atomic groups is compared to that between a pair of \#3 atomic groups. The PMF between a pair of \#2 atomic groups is compared to that between a pair of \#8 atomic groups. The PMF between \#1 atomic group and \#8 atomic group is compared to that between \#3 atomic group and \#2 atomic group. The results show that after the subtraction of the Coulombic part, the resulting van der Waals PMF is almost the same among the atomic groups mapped to the same coarse-grained beads, justifying the usage of only 5 types of coarse-grained beads for the nonbonded van der Waals interactions as discussed in the main text.


\noindent \textbf{S3: Comparing Diffusion and Size Distribution of Chains}

\begin{figure}[htb]
  \centering
  \includegraphics[width = 0.35 \textwidth]{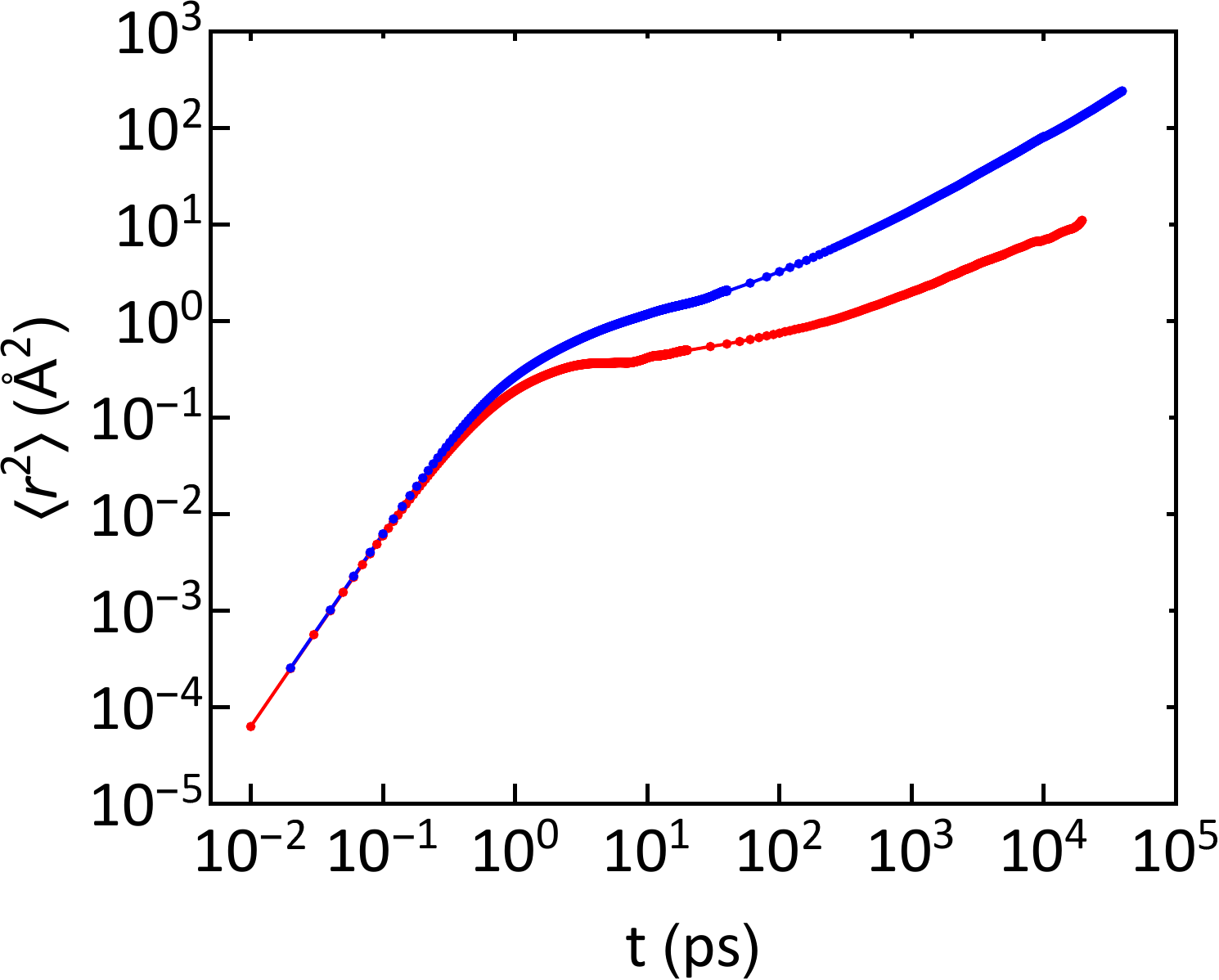}
  \caption{Mean square displacement, $\langle r^2\rangle$, of the center of mass of chains from the all-atom model (red) and the $\text{CG}_\alpha^n$ model (blue) as a function of time $t$.}
  \label{fg:diffusion_comparison}
\end{figure}

In Fig.~\ref{fg:diffusion_comparison} we compare the mean square displacement, $\langle r^2\rangle$, of the center of mass of branched polyetherimide chains in a melt state at 700 K from the all-atom model and the $\text{CG}_\alpha^n$ coarse-grained model. In the all-atom system, there are 64 chains while in the coarse-grained model, there are 1000 chains. At very short times, the motion is ballistic. The chain motion becomes diffusive at long times. As expected, the diffusion is faster in the coarse-grained model as the coarse-grained bead-bead potentials are smoother than the interatomic ones. The diffusion coefficient $D$ can be computed as
\begin{equation}
D= \lim_{t\rightarrow \infty} \frac{\langle r^2 \rangle} {6t}~.
\end{equation}
The value of $D$ at 700 K is about $1.0\times 10^{-11}~\text{m}^2/\text{s}$ in the $\text{CG}_\alpha^n$ model and about $0.092\times 10^{-11}~\text{m}^2/\text{s}$ in the all-atom model. The chain diffusion is therefore about 11 times faster in the coarse-grained model.

\begin{figure}[htb]
  \centering
  \includegraphics[width = 0.35 \textwidth]{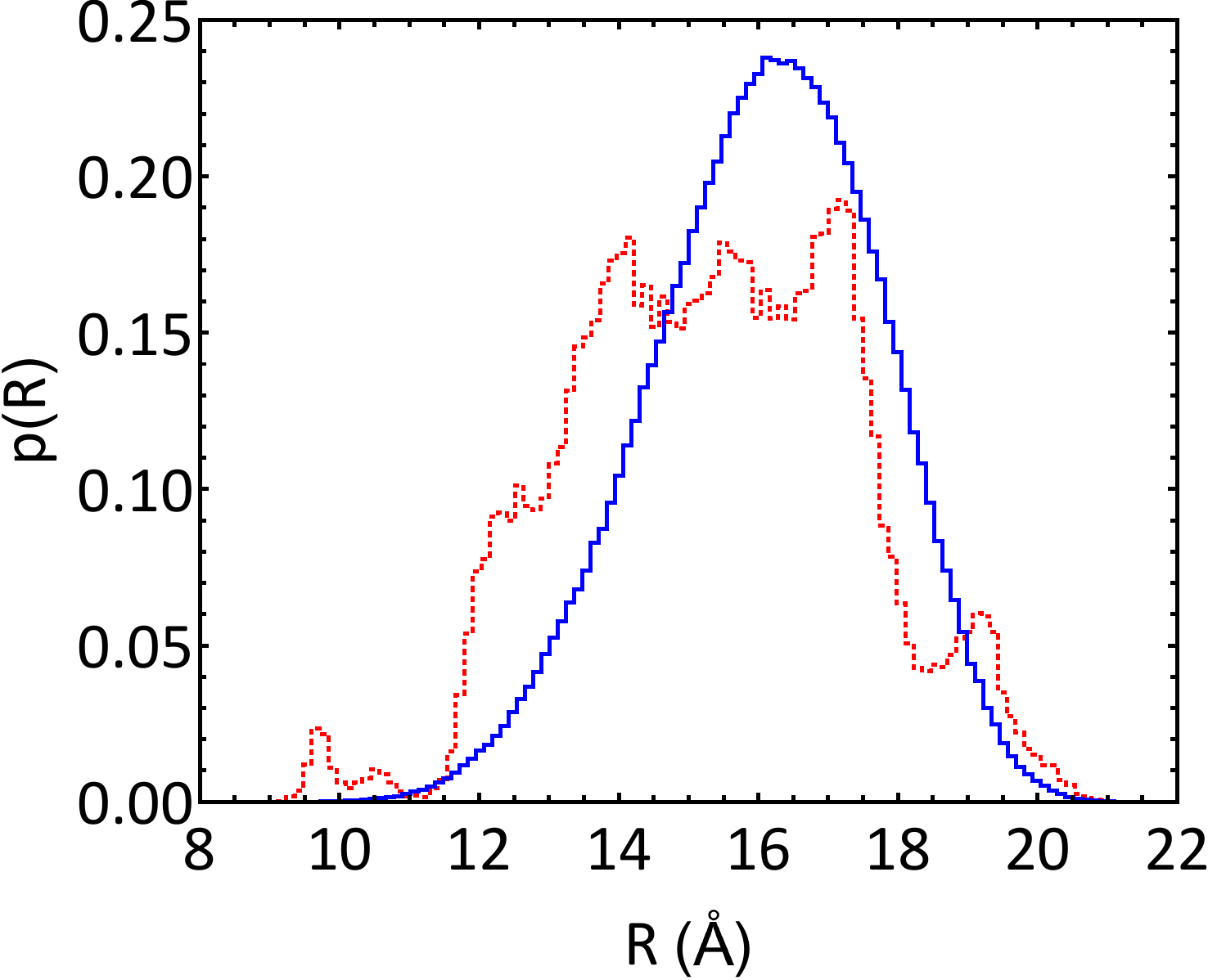}
  \caption{Probability density $p(R)$ of chain size $R$ from the all-atom (red) and $\text{CG}_\alpha^n$ (blue) models. The value of $R$ is plotted in increments of 0.1214~\AA~and 0.1172~\AA~for the all-atom data and the coarse-grained data, respectively.}
  \label{fg:chain_dist}
\end{figure}

We also compute the radius of gyration, $R$, of the branched polyetherimide chains in a melt state at 700 K from the all-atom model and the $\text{CG}_\alpha^n$ coarse-grained model. The probability distribution $p(R)$ is shown in Fig.~\ref{fg:chain_dist}, which shows a reasonable agreement between the two models. The all-atom system only consists of 64 chains, resulting in the relatively poor statistics of the $p(R)$ curve for that system in Fig.~\ref{fg:chain_dist}. The average value of $R$ is 15.3~\AA~and 16.0~\AA~for the all-atom and $\text{CG}_\alpha^n$ model, respectively.

\noindent \textbf{S4: Coarse-Grained Force Field of the Branched Polyetherimide}

\noindent The following two files are available upon request: (1) The final coarse-grained force field of the branched polyetherimide, which can be read directly by LAMMPS; (2) A sample input script, which can be added to a LAMMPS script to read in this coarse-grained force field.

\end{document}